\documentclass[10pt,a4paper]{article}
\usepackage[utf8]{inputenc}
\usepackage{amsmath}
\usepackage{amsfonts}
\usepackage{amssymb}
\usepackage{amsthm}
\usepackage{graphicx}
\usepackage{lmodern}
\usepackage{mathtools}
\usepackage{color}
\usepackage{subcaption}

\usepackage[left=2cm,right=2cm,top=2cm,bottom=2cm]{geometry}

\newtheorem{theorem}{Theorem}[section]

\newtheorem{lemma}[theorem]{Lemma}

\newenvironment{proof1}{\paragraph{Proof:}}{\hfill$\square$}
\theoremstyle{remark}
\newtheorem{remark}{Remark}[section]
\newcommand\keywords[1]{%
    \begingroup
    \let\and\\
    \par
    \noindent\emph{Keywords:} #1\par
    \endgroup
}

\newcommand\msc[1]{%
    \begingroup
    \let\and\\
    \par
    \noindent\emph{2010 MSC:} #1\par
    \endgroup
}

\makeatother

\author{P. Amster$^{1,2}$ and A.P. Mogni$^1$}
\title{Adapting the CVA model to Leland's framework}
\begin{document}

\date{}
\maketitle
\begin{center}
$^1$ Departamento de Matem\'atica, \\
Facultad de Ciencias Exactas y Naturales\\
Universidad de Buenos Aires
and \\
$^2$ IMAS - CONICET\\
Ciudad Universitaria, Pabell\'on I,
1428 Buenos Aires, Argentina \\
{\sl E-mails}: pamster@dm.uba.ar --- amogni@dm.uba.ar
\end{center}

\date{}

\begin{abstract}

\noindent We consider the framework proposed by Burgard and Kjaer in \cite{burgard2011partial} that derives the PDE which governs the price of an option including bilateral counterparty risk and funding. We extend this work by relaxing the assumption of absence of transaction costs in the hedging portfolio by proposing a cost proportional to the amount of assets traded and the traded price. After deriving the nonlinear PDE, we prove the existence of a solution for the corresponding initial-boundary value problem. Moreover, we develop a numerical scheme that allows to find the solution of the PDE by setting different values for each parameter of the model. To understand the impact of each variable within the model, we analyze the Greeks of the option and the sensitivity of the price to changes in all the risk factors. \\


\msc{35K20, 35K55, 91G20, 91G60.}

\end{abstract}

\maketitle

\section{Introduction}

\noindent Under the Black-Scholes pricing framework \cite{black1973pricing}, the price of an option is derived by constructing a hedging portfolio consisting of a certain amount of the underlying asset and a money-market account. This methodology relies on a list of different assumptions that are set to simplify the model. For example, constant values of volatility and interest rates, the non-existence of dividend yields, the efficiency of the markets and the non-existence of transaction costs, among others. Nonetheless, the probability of default of both the issuer of the option and the counterparty are not being considered when constructing the hedging portfolio. Therefore, it is expected that the option price obtained by the standard modeling approach will be different when compared with the actual price of the contract.\\

\noindent As explained in \cite{gregory2009being}, the counterparty credit risk is the risk that a counterparty in a financial contract defaults prior to the expiration of the contract and fails to make future payments. To include this probability in the pricing of a contract, the Credit Valuation Adjustment (CVA) is used. As described in \cite{green2015xva}, CVA is defined as the difference between the price of the instrument including credit risk and the price where the counterparty of the transaction is considered free of risk. By definition, the CVA will be always positive if only the counterparty risk is considered. When the issuer credit riskiness is also taken into account, the Debit Valuation Adjustment (DVA) is included into the formula. The DVA acts oppositely as the CVA by adding value to the option when the issuer risk increases. If we set $V_{BS}$ as the Black-Scholes option price default free and $V$ as the adjusted option price we get

\begin{align*}
V  \, = \, V_{BS} \, - \, CVA \, + \, DVA
\end{align*}

\noindent where $CVA$ is a cost and $DVA$ is a benefit.\\

\noindent As mentioned in \cite{green2015xva}, before 2008 crisis, CVA was commonly calculated and charged only by tier one banks by choosing either unilateral or bilateral models. After 2008 crisis, not only bilateral CVA started to be widely used but also a different set of value adjustments began to be applied such as the Funding Valuation Adjustment or FVA (including costs of of funding) and the Capital Valuation Adjustment or KVA (including cost of capital).\\

\noindent Several works have been developed within the family of value adjustments for different OTC derivatives. In \cite{alavian2008counterparty}, the authors derive the CVA by decomposing the portfolio's value into a set of binary states: positive or negative cash flows directions, two possible default states and to receive or not the recovery rate. In \cite{brigo2009counterparty}, the authors study a CDS pricing model which includes the probability of default of the counterparty. 
Credit spread volatilities are considered by assuming default intensities following a CIR dynamic with a correlation parameter within. The authors in \cite{brigo2008bilateral} generalize these model in two ways. First, by allowing correlation not only between default times but also with the underlying portfolio risk factors (interest rates). Second, by considering the probability of default of the issuer.\\

\noindent These previous works are then expanded by \cite{brigo2011arbitrage} in order to apply that methodology on IR Swaps and IR Exotics. DVA and FVA are included in the pricing framework by \cite{pallavicini2012funding} where a recursive formula is obtained after discounting all the cash flows occurring after the trading position is entered. These cash flows include not only the product cash flows (coupons, dividends, etc) but also cash flows required by collateral margining, funding and investing procedures and default events.\\

\noindent Finally, two works try to propose a general theory. In \cite{burgard2011partial} the authors analyze the bilateral counterparty risk (CVA and DVA) with funding costs (FVA) by constructing a hedging portfolio composed by the underlying asset, a risk-free zero-coupon bond and two default risky zero-coupon bond of both the issuer and the counterparty. A PDE is formalized by applying the standard self-financing assumption. In \cite{crepey2011bsde} (and further in \cite{crepey2015bilateral1,crepey2015bilateral2}), the author applies a risk-neutral pricing approach under funding constraints in order to obtain a reduced-form backward stochastic differential equation (BSDE) approach to the problem of hedging and pricing the CVA.\\

\noindent The second assumption covered in our model is the (non) existence of transaction costs. Following Leland's approach \cite{leland1985option}, transaction costs can be included in the pricing methodology by applying a discrete-time replicating strategy. A nonlinear partial differential equation is obtained for the option price , which is denoted by $V\left(S,t\right)$; namely,

\begin{align*}
\frac{\partial V}{\partial t}+\frac{1}{2} \hat{\sigma} \left( S \frac{\partial^2 V}{\partial S^2} \right)^2 S^2  \frac{\partial^2 V}{\partial S^2} +rS\frac{\partial V}{\partial S}-rV=0,
\end{align*}

\noindent where $\hat{\sigma}$ is defined based upon the transaction costs function.  This approach was then continued and improved by \cite{boyle1992option} and \cite{hoggard1994hedging}.\\

\noindent Different choices of transaction costs functions lead to variations on the nonlinear term of the partial differential equation. In \cite{amster2005black}, the authors propose a non-increasing linear function and find solutions for the stationary problem. In \cite{vsevvcovivc2016analysis}, the concept of transaction costs function is generalized and the so-called \emph{mean value modification of the transaction costs function} is developed. This transformation allows the authors to formulate a general one-dimensional Black-Scholes equation by solving the equivalent quasilinear Gamma equation. Other authors \cite{sengupta2014option,florescu2014option} also find solutions to the problem with constant transaction costs and relaxing the assumptions of constant volatility and interest rate.\\

\noindent The main distinctive aspect in the above-cited works is that they all consider only one asset within the partial differential equation. In \cite{zakamouline2008hedging} and \cite{zakamulin2008option}, the author generalizes the Leland approach to cover different types of multi-asset options, developing the nonlinear partial differential equation and solving numerically a list of examples. In \cite{amster2017pricing} the authors combine a multidimensional approach with a general transaction cost function to derive a the dynamics of the option price under a fully nonlinear PDE.\\

\noindent In this work we adapt the work of Burgard and Kjaer \cite{burgard2011partial} in order to include the transaction costs generated by trading the underlying assets and both the issuer and counterparty bonds. We propose an initial constant transaction cost function proportional to the amount of assets traded. As a consequence, we derive a nonlinear PDE that extends the results found in \cite{burgard2011partial} and prove the existence of a solution by applying the Schauder Fixed-Point theorem. In the second part of the work, we develop a numerical approach to solve the PDE by considering a non-uniform grid on the spatial variable. The main greeks of the option (Delta, Gamma, Rho and Vega) are calculated and analyzed to understand how both the value adjustments and transaction costs affect the behavior of the option price. Nonetheless, a sensitivity analysis on the remaining parameters (hazard rate, recovery rates, etc) is performed to complete the study of the option price dynamics.\\

\noindent The structure of the paper is as follows. In Section 2 we propose the market model that leads to the nonlinear dynamic of the option price. In Section 3 we apply the Schauder Fixed-Point theorem to derive the existence of solution for the original problem. Finally, in Section 4, the numerical framework is developed and different results are obtained to understand how the parameters affect the option price.\\

\section{Market Model}

\noindent The original paper of \cite{burgard2011partial} derives the PDE for the value of a financial derivative considering bilateral counterparty risk and funding costs. For this purpose they propose an economy consisting of a risk-free zero-coupon bond, two default risky zero-coupon bond with zero recovery of parties B and C and a spot asset with no default risk. B will refer to the seller and C to the counterparty. Notation will be followed from the original work \cite{burgard2011partial}.\\

\noindent The dynamics of the four tradable assets under the historical probability measure are defined as follows:

\[
\begin{cases}
dP_R &= P_R \, r \, dt,\\
dP_B &= P_B \, r_B \, dt \, - \, P_B \, dJ_B,\\
dP_C &= P_C \, r_C \, dt \, - \, P_C \, dJ_C,\\
dS &= \mu \, S \, dt \, + \, S \, dW_t. 
\end{cases}
\]

\noindent The default risky zero-coupon bonds are modeled by considering both $r_B$ and $r_C$ interest rates and $J_B$ and $J_C$ the two independent point processes that jump from $0$ to $1$ on default of B and C respectively. The default risk-free zero-coupon bond is a deterministic process with drift equal to $r$ and the spot asset is modeled following a geometric brownian motion with drift $\mu$ and volatility $\sigma$. Throughout this work the parameters $r,r_B,r_C,\mu$ and $\sigma$ are positive and constant. Also we will use the following notation:

\begin{align*}
x^{+} \, &= \, \max \left(x,0\right)\\
x^{-} \, &= \, \min \left(x,0\right).
\end{align*}

\noindent To derive the price of option $\hat{V}$, we adapt the standard Black-Scholes framework \cite{black1973pricing} applied in \cite{burgard2011partial} by considering the Leland's approach \cite{leland1985option}. Hence, we create a self-financing portfolio covering all the underlying risk factors that hedges the option. Let $\Pi\left(t\right)$ be the seller's portfolio which consists of $\delta\left(t\right)$ units of $S\left(t\right)$, $\alpha_B\left(t\right)$ units of $P_B\left(t\right)$, $\alpha_C\left(t\right)$ units of $P_C\left(t\right)$ and $\beta\left(t\right)$ units of cash. For hedging purposes we set $\Pi\left(t\right)+\hat{V}\left(t\right)=0$ and

\begin{align}
-\hat{V}\left(t\right)=\Pi\left(t\right)=\delta\left(t\right) \, S\left(t\right) \, + \, \alpha_B\left(t\right) \, P_B\left(t\right) \, + \, \alpha_C\left(t\right) \, P_C\left(t\right) \, + \, \beta\left(t\right). \label{rep_port}
\end{align}

\noindent We define the transaction costs function for both default risky bonds $P_B$ and $P_C$ and the spot asset $S$ as follows:

\[
\begin{cases}
TC_B \left(t,P_B\right) &= C_B \, \vert \alpha_B\left(t\right) \vert \, P_B\left(t\right),\\
TC_C \left(t,P_C\right)&= C_C \, \vert \alpha_C\left(t\right) \vert \, P_C\left(t\right),\\
TC_S \left(t,S\right)&= C_S \, \vert \delta\left(t\right) \vert \, S\left(t\right)
\end{cases}
\]

\noindent where $C_B,C_C$ and $C_S$ are positive constants. This definition of transaction costs is the standard approach applied initially in \cite{leland1985option} and is the initial step to creating more complex dynamics. In this case, the costs are defined to be proportional to the amount of assets traded multiplied by the price of each asset. For the purpose of enhancing clarity, we drop the dependencies on every function.\\

\noindent By forcing the portfolio to be self-financing, we find that

\begin{align}
-d\hat{V}=\delta \, dS \, + \, \alpha_B \, dP_B \, + \, \alpha_C \, dP_C \, + \, d\beta. \label{reportfolio}
\end{align} 

\noindent where $d\beta$ is decomposed as $d\beta=d\beta_S+d\beta_F+d\beta_C$ corresponding to the variations in the cash position due to each of the three assets. In this step we consider the effect of the transaction cost in the hedging strategy. On each time step, there would be a decrease in the cash account because of the cost of buying or selling a different amount of assets. Hence, the original calculations of \cite{burgard2011partial} are modified as follows:

\begin{itemize}
\item The share position provides a dividend income, a financing cost and a transaction cost. The variation in the position is found to be
\begin{align}
d\beta_S=\delta \, \gamma_S \, S \, dt - \delta \, q_S \, S \, dt - dTC_S \label{betaS}
\end{align}
\item After the own bonds are purchased, if any surplus in cash is available, it must earn the free-risk-rate $r$. If borrowing money, the seller needs to pay the rate $r_F$. In this case, transaction costs appear when calculating the surplus after the own bonds purchasing. The variation in this position is determined by
\begin{align}
d\beta_F &= r\, \left( -\hat{V}-\alpha_B \, P_B - TC_B \right)^{+} \, dt + r_F \left( -\hat{V}-\alpha_B \, P_B - TC_B \right)^{-} \, dt \nonumber \\
&=r\, \left( -\hat{V}-\alpha_B \, P_B - TC_B \right) \, dt + s_F \left( -\hat{V}-\alpha_B \, P_B - TC_B \right)^{-} \, dt \label{betaF}
\end{align} 
\noindent where $s_F=r_F-r$ is the funding spread.\\
\item Finally, a financing cost due to short-selling the counterparty bond and its related transaction costs are considered for calculating the variation in the cash counterparty position as follows:
\begin{align}
d\beta_C= - \alpha_C \, r \, P_C \, dt - dTC_C. \label{betaC}
\end{align}
\end{itemize}

\noindent By applying equations \eqref{betaS}, \eqref{betaF} and \eqref{betaC} in \eqref{reportfolio}, we obtain

\begin{align}
-d\hat{V}&= \delta \, dS + \alpha_B \, P_B \left(r_B \, dt -dJ_B \right)+ \alpha_C \, P_C \left(r_C \, dt -dJ_C \right)- dTC_C-dTC_S \, +\\
 &\left[\, \left( -\hat{V}-\alpha_B \, P_B - TC_B \right)^{+} + r_F \left( -\hat{V}-\alpha_B \, P_B - TC_B \right)^{-} + \delta \left(\gamma_S - q_S \right) \, S - \alpha_C \, r \, P_C \ \right] dt  \nonumber\\
 &=\left[-r\hat{V}+s_F \left( -\hat{V}-\alpha_B \, P_B - TC_B \right)^{-} + \left(r_B - r \right) \alpha_B \, P_B + \left(r_C - r \right) \alpha_C \, P_C - r \, TC_B + \delta \, \left(\gamma_S - q_S \right)\right] dt \nonumber\\
 &- dTC_S - dTC_C + \delta \, dS - \alpha_B \, P_B \, dJ_B - \alpha_C \, P_C \, dJ_C. \label{portfolio2}
\end{align}

\noindent Also, the Ito lemma adapted for jump processes provides another equivalent calculation of the variation of the process $\hat{V}$:

\begin{align}
d\hat{V}=\frac{\partial \hat{V}}{\partial t} \, dt + \frac{\partial \hat{V}}{\partial S} \, dS + \frac{1}{2}\sigma^2 S^2 \frac{\partial^2 \hat{V}}{\partial S^2} \, dt + \bigtriangleup \hat{V}_B \, dJ_B + \bigtriangleup \hat{V}_C \, dJ_C \label{ito}
\end{align}

\noindent where $\bigtriangleup \hat{V}_B$ and $\bigtriangleup \hat{V}_C$ are calculated based on default conditions. In \cite{burgard2011partial} it is showed that

\begin{align*}
\bigtriangleup \hat{V}_B &=\hat{V}\left(t,S,1,0\right)-\hat{V}\left(t,S,0,0\right)= - \hat{V} + \hat{V}^{+}+ \, R_B \, \hat{V}^{-}\\
\bigtriangleup \hat{V}_C &=\hat{V}\left(t,S,0,1\right)-\hat{V}\left(t,S,0,0\right)= - \hat{V} + R_C \, \hat{V}^{+}+ \, \hat{V}^{-}
\end{align*}

\noindent Hence, by adding \eqref{portfolio2} and \eqref{ito}, and in order to hedge the risks related to the corporate bonds and the spot asset we find that

\begin{align*}
\delta &= - \frac{\partial \hat{V}}{\partial S}\\
\alpha_B &= \frac{- \hat{V} + \hat{V}^{+}+ \, R_B \, \hat{V}^{-}}{P_B}\\
\alpha_C &= \frac{- \hat{V} + R_C \, \hat{V}^{+}+ \, \hat{V}^{-}}{P_C}
\end{align*}

\noindent and

\begin{align}
0=&\left[-r\hat{V}+s_F \left( -\hat{V}-\alpha_B \, P_B - TC_B \right)^{-} + \left(r_B - r \right) \alpha_B \, P_B + \left(r_C - r \right) \alpha_C \, P_C - r \, TC_B + \delta \, S \, \left(\gamma_S - q_S \right) + \right. \nonumber \\
&\left. \frac{\partial \hat{V}}{\partial t} + \frac{1}{2}\sigma^2 S^2 \frac{\partial^2 \hat{V}}{\partial S^2}\right] \, dt - dTC_S - dTC_C  \label{eq}
\end{align}

\noindent By recalling the definition of the transaction costs, we can compute $dTC_S$ and $dTC_C$. Then, for the calculation of the transaction costs of the spot asset, we recall the value of $\delta$ and note that

\begin{align}
dTC_S = C_S \, \lvert d\delta \rvert \, S \,  \sim C_S \, \left\lvert  \frac{\partial^2 \hat{V}}{\partial S^2} \right\rvert \, \sigma \, S^2 \sqrt{dt} \, \sqrt{\frac{2}{\pi}} \label{dTC_S}
\end{align}

\noindent where the approximation is made by taking the expected value of $\lvert d\delta \rvert$ and the lowest order $\mathcal{O}\left(\sqrt{\Delta t} \right)$ as follows 

\begin{align*}
E\left(\left\lvert d\delta \right\rvert \right)= E\left( \, \left\lvert \frac{\partial^2 \hat{V}}{\partial S^2} \right\rvert \, dS \right)= \left\lvert \frac{\partial^2 \hat{V}}{\partial S^2} \right\rvert \sigma \, S \, \sqrt{dt} \, E\left[\Phi \right]\\
\end{align*}

\noindent and setting $\Phi$ as a standard normal random variable.\\

\noindent The variation of the transaction costs of the counterparty bond position is computed by applying the same rationale as before but over $\lvert d\alpha_C \rvert$ in this case. Then,

\begin{align}
dTC_C = C_C \, \lvert d\alpha_C \rvert \, P_C \,  \sim C_C \, \left\lvert R_C \, \frac{\partial \hat{V}}{\partial S}^{+} + \frac{\partial \hat{V}}{\partial S}^{-} -\frac{\partial \hat{V}}{\partial S} \right\rvert \, \sigma \, S \sqrt{dt} \, \sqrt{\frac{2}{\pi}} \label{dTC_C}
\end{align}

\noindent where in this occasion the approximation is obtained by taking the expected value of $\lvert d\alpha_C \rvert$ as follows 

\begin{align*}
E\left(\left\lvert d\alpha_C \right\rvert \right)= E\left( \, \, \left\lvert R_C \, \frac{\partial \hat{V}}{\partial S}^{+} + \frac{\partial \hat{V}}{\partial S}^{-} -\frac{\partial \hat{V}}{\partial S} \right\rvert \, dS \right)= \left\lvert R_C \, \frac{\partial \hat{V}}{\partial S}^{+} + \frac{\partial \hat{V}}{\partial S}^{-} -\frac{\partial \hat{V}}{\partial S} \right\rvert \sigma \, S \, \sqrt{dt} \, E\left[\Phi \right]\\
\end{align*}

\noindent and setting again $\Phi$ as a standard normal random variable.\\

\noindent By recalling \eqref{dTC_S} and \eqref{dTC_C} and applying those computations in \eqref{eq}, we obtain the following nonlinear parabolic partial derivative equation

\begin{align}
0=&\left[-r\hat{V}+s_F \left( -\hat{V}-\alpha_B \, P_B - TC_B \right)^{-} + \left(r_B - r \right) \alpha_B \, P_B + \left(r_C - r \right) \alpha_C \, P_C - r \, TC_B + \delta \, S \, \left(\gamma_S - q_S \right) + \right. \nonumber \\
&\left. \frac{\partial \hat{V}}{\partial t} + \frac{1}{2}\sigma^2 S^2 \frac{\partial^2 \hat{V}}{\partial S^2}\right] \, dt - C_S \, \left\lvert  \frac{\partial^2 \hat{V}}{\partial S^2} \right\rvert \, \sigma \, S^2 \sqrt{dt} \, \sqrt{\frac{2}{\pi}} - C_C \, \left\lvert R_C \, \frac{\partial \hat{V}}{\partial S}^{+} + \frac{\partial \hat{V}}{\partial S}^{-} -\frac{\partial \hat{V}}{\partial S} \right\rvert \, \sigma \, S \sqrt{dt} \, \sqrt{\frac{2}{\pi}} \nonumber \\
=& \left[-r\hat{V}+s_F \left( -\hat{V}-\alpha_B \, P_B - C_B \, \left\lvert - \hat{V} + \hat{V}^{+}+ \, R_B \, \hat{V}^{-} \right\rvert \right)^{-} + \left(r_B - r \right) \alpha_B \, P_B + \left(r_C - r \right) \alpha_C \, P_C \right. \nonumber \\
&\left. - r \, \left(C_B \, \left\lvert - \hat{V} + \hat{V}^{+}+ \, R_B \, \hat{V}^{-} \right\rvert \right) - \frac{\partial \hat{V}}{\partial S} \, S \, \left(\gamma_S - q_S \right) + \frac{\partial \hat{V}}{\partial t} + \frac{1}{2}\sigma^2 S^2 \frac{\partial^2 \hat{V}}{\partial S^2}\right] \, dt - C_S \, \left\lvert  \frac{\partial^2 \hat{V}}{\partial S^2} \right\rvert \, \sigma \, S^2 \sqrt{dt} \, \sqrt{\frac{2}{\pi}} \nonumber \\
&- C_C \, \left\lvert R_C \, \frac{\partial \hat{V}}{\partial S}^{+} + \frac{\partial \hat{V}}{\partial S}^{-} -\frac{\partial \hat{V}}{\partial S} \right\rvert \, \sigma \, S \sqrt{dt} \, \sqrt{\frac{2}{\pi}} \label{eq2}
\end{align}

\noindent By setting $\lambda_B=r_B-r$, $\lambda_C=r_C-r$ and applying the definitions of $\alpha_B$ and $\alpha_C$, \eqref{eq2} becomes 

\begin{align}
0=& \left[-r\hat{V}+s_F \left( -\hat{V}^{+}- \, R_B \, \hat{V}^{-}  - C_B \, \left\lvert - \hat{V} + \hat{V}^{+}+ \, R_B \, \hat{V}^{-} \right\rvert \right)^{-} + \lambda_B \left(- \hat{V} + \hat{V}^{+}+ \, R_B \, \hat{V}^{-} \right) \right. \nonumber \\
+& \left. \lambda_C \left(- \hat{V} + R_C \, \hat{V}^{+}+ \, \hat{V}^{-} \right) - r \, \left(C_B \, \left\lvert - \hat{V} + \hat{V}^{+}+ \, R_B \, \hat{V}^{-} \right\rvert \, \right) - \frac{\partial \hat{V}}{\partial S} \, S \, \left(\gamma_S - q_S \right) +  \frac{\partial \hat{V}}{\partial t} + \frac{1}{2}\sigma^2 S^2 \frac{\partial^2 \hat{V}}{\partial S^2}\right] \, dt \nonumber \\
-&  C_S \, \left\lvert  \frac{\partial^2 \hat{V}}{\partial S^2} \right\rvert \, \sigma \, S^2 \sqrt{dt} \, \sqrt{\frac{2}{\pi}}- C_C \, \left\lvert R_C \, \frac{\partial \hat{V}}{\partial S}^{+} + \frac{\partial \hat{V}}{\partial S}^{-} -\frac{\partial \hat{V}}{\partial S} \right\rvert \, \sigma \, S \sqrt{dt} \, \sqrt{\frac{2}{\pi}} \label{eq3}
\end{align}

\noindent The absolute value that involves the transaction costs due to the own bonds purchase can be reduced by noting that when $\hat{V}\geq 0$, its value is $0$ and when $\hat{V}< 0$ it is equal to $\left(R_B - 1\right) \, \hat{V}$. Hence, 

\begin{align*}
\left\lvert - \hat{V} + \hat{V}^{+}+ \, R_B \, \hat{V}^{-} \right\rvert = \left(R_B - 1 \right) \, \hat{V}^{-}
\end{align*}

\noindent Using this reduction in \eqref{eq3}, we get:

\begin{align}
\left( -\hat{V}^{+}- \, R_B \, \hat{V}^{-}  - C_B \, \left\lvert - \hat{V} + \hat{V}^{+}+ \, R_B \, \hat{V}^{-} \right\rvert \right)^{-} &= \left( -\hat{V}^{+}- \, R_B \, \hat{V}^{-}  - C_B \, \left(R_B - 1 \right) \, \hat{V}^{-} \right)^{-} \nonumber \\
&= \left( -\hat{V}^{+} - \hat{V}^{-} \left[R_B - C_B \, \left(R_B - 1 \right) \right]\right)^{-} \nonumber\\
&= \begin{cases}
- \hat{V} \quad \text{if} \quad \hat{V}\geq 0\\
\quad 0 \quad \text{if} \quad \hat{V}< 0\\
\end{cases} \nonumber \\
&= \, - \hat{V}^{+}. \label{tcb_abs}
\end{align} 

\noindent Thus, by implementing \eqref{tcb_abs} in \eqref{eq3}, 

\begin{align}
0=& -r\hat{V}-s_F \, \hat{V}^{+} + \lambda_B \left(- \hat{V} + \hat{V}^{+}+ \, R_B \, \hat{V}^{-} \right) +  \lambda_C \left(- \hat{V} + R_C \, \hat{V}^{+}+ \, \hat{V}^{-} \right) - r \, C_B \, \left(R_B - 1 \right) \, \hat{V}^{-}  \nonumber \\
-& \frac{\partial \hat{V}}{\partial S} \, S \, \left(\gamma_S - q_S \right) +  \frac{\partial \hat{V}}{\partial t} + \frac{1}{2}\sigma^2 S^2 \frac{\partial^2 \hat{V}}{\partial S^2} - \sigma \, S^2  \, \sqrt{\frac{2}{\pi \, dt}} \left( C_S \, \left\lvert  \frac{\partial^2 \hat{V}}{\partial S^2} \right\rvert \, + S^{-1} \, C_C \, \left\lvert R_C \, \frac{\partial \hat{V}}{\partial S}^{+} + \frac{\partial \hat{V}}{\partial S}^{-} -\frac{\partial \hat{V}}{\partial S} \right\rvert \, \right). \label{eq4}
\end{align}

\noindent If we introduce the parabolic operator $\mathcal{A}_t$ as 

\begin{align*}
\mathcal{A}_t \equiv \frac{1}{2}\sigma^2 S^2 \frac{\partial^2 \hat{V}}{\partial S^2} + \frac{\partial \hat{V}}{\partial S} \, S \, \left(q_S - \gamma_S \right)
\end{align*}

\noindent then it follows that $\hat{V}$ is the solution of 

\begin{align}
0=& \, \frac{\partial \hat{V}}{\partial t} + \mathcal{A}_t \hat{V} -  \left(\lambda_B + \lambda_C + r \right) \, \hat{V} +  \left(\lambda_B + \lambda_C \, R_C - s_F \right) \, \hat{V}^{+} + \left(\lambda_B \, R_B + \lambda_C - r \, \left(R_B - 1 \right) \, C_B \right)  \, \hat{V}^{-}  \nonumber \\
-& \, \sigma \, S^2  \, \sqrt{\frac{2}{\pi \, dt}} \left( C_S \, \left\lvert  \frac{\partial^2 \hat{V}}{\partial S^2} \right\rvert \, + S^{-1} \, C_C \, \left\lvert R_C \, \frac{\partial \hat{V}}{\partial S}^{+} + \frac{\partial \hat{V}}{\partial S}^{-} -\frac{\partial \hat{V}}{\partial S} \right\rvert \, \right) \label{eq5}
\end{align}

\noindent Looking forward to compare \eqref{eq5} with equation 26 from \cite{burgard2011partial}, we rearrange the terms that involve $\hat{V}$, $\hat{V}^{+}$ and $\hat{V}^{-}$ and obtain the following nonlinear parabolic PDE

\begin{align}
 \frac{\partial \hat{V}}{\partial t} + \mathcal{A}_t \hat{V} -  r \, \hat{V} =& \, s_F \, \hat{V}^{+} + \lambda_C \, \left(1-R_C \right) \, \hat{V}^{+} + \lambda_B \, \left(1-R_B \right) \, \hat{V}^{-} - r \, \left(1-R_B \right) \, C_B \, \hat{V}^{-}  \nonumber \\
+& \, \sigma \, S^2  \, \sqrt{\frac{2}{\pi \, dt}}  C_S \, \left\lvert  \frac{\partial^2 \hat{V}}{\partial S^2} \right\rvert \, + \sigma \, S  \, \sqrt{\frac{2}{\pi \, dt}} C_C \, \left\lvert R_C \, \frac{\partial \hat{V}}{\partial S}^{+} + \frac{\partial \hat{V}}{\partial S}^{-} -\frac{\partial \hat{V}}{\partial S} \right\rvert.  \label{eq6}
\end{align}

\noindent The first three terms on the right hand side of \eqref{eq6} are equal to the nonlinear terms of the original model. The inclusion of the transaction costs in the hedging strategy brings to the model three new terms:

\begin{itemize}
\item The fourth term in the right hand side of \eqref{eq6} corresponds to the amount of cash that is not invested at $r$ rate when considering the surplus held by the seller after the purchase of its own bonds as it is shown in \eqref{betaF}.\\
\item The fifth term is the effect of the transaction costs due to buying or selling $\delta$ assets of $S$. It shall be noted that the term is equal to the corresponding one  in Leland's standard approach.\\
\item The sixth term is the effect of the transaction costs due to shorting the counterparty bond.  
\end{itemize}
 
\noindent Comparing \eqref{eq5} with Leland's notation, we can define the \emph{modified volatility} as

\begin{align}
\hat{\sigma}^2= \sigma^2 \, \left(1 -  \sqrt{\frac{2}{\pi \, dt}} \,  \frac{C_S}{\sigma} \, \text{sgn}\left( \frac{\partial^2 \hat{V}}{\partial S^2} \right) \right) \label{mod_vol}
\end{align}

\noindent and noting that

\begin{align}
\left\lvert R_C \, \frac{\partial \hat{V}}{\partial S}^{+} + \frac{\partial \hat{V}}{\partial S}^{-} -\frac{\partial \hat{V}}{\partial S} \right\rvert &= \begin{cases}
\left\lvert \left(1-R_C \right) \, \frac{\partial \hat{V}}{\partial S} \right\rvert \quad &\text{if} \quad \hat{V}\geq 0\\
\quad 0 \quad &\text{if} \quad \hat{V}< 0\\
\end{cases} \nonumber \\
&= \, \left(1-R_C \right) \, \left\lvert \, \frac{\partial \hat{V}}{\partial S}^{+} \right\rvert
\end{align}

\noindent we obtain the following differential equation

\begin{align}
\frac{\partial \hat{V}}{\partial t} + \frac{1}{2} \, \hat{\sigma}^2 \, S^2 \, \frac{\partial^2 \hat{V}}{\partial S^2} + \frac{\partial \hat{V}}{\partial S} \, S \, \left(q_S - \gamma_S \right) -  r \, \hat{V} =& \, s_F \, \hat{V}^{+} + \lambda_C \, \left(1-R_C \right) \, \hat{V}^{+} + \lambda_B \, \left(1-R_B \right) \, \hat{V}^{-} - r \, \left(1-R_B \right) \, C_B \, \hat{V}^{-}  \nonumber \\
+& \sigma \, S  \, \sqrt{\frac{2}{\pi \, dt}} C_C \, \left(1-R_C \right) \, \left\lvert \, \frac{\partial \hat{V}}{\partial S}^{+} \right\rvert \label{eq7}
\end{align}

\begin{remark}
The left-hand side of equation \eqref{eq7} is effectively a Black-Scholes operator with a volatility parameter $\hat{\sigma}$, a dividend yield $\gamma_S$ and a financing cost (different to the risk-free interest rate) $q_S$. The right-hand side of the equation contains the nonlinear terms that arises from considering the existence of transaction costs and default risk. The inclusion of these 'extra' costs can be thought as a perturbation to the original model. By assessing the magnitude of the parameters of each term, it can be noted that they are indeed small. Hazard rates, recovery rates and interest rates are always below $1$ and the transaction costs per unit of asset can be modeled between $0.025$ to $0.04$ as it is done in \cite{leland1985option}.
\end{remark}

\section{Existence}
\subsection{Preliminaries}

\noindent For the purpose of finding the unique solution of the CVA problem with transaction costs, we set the mathematical framework in which we will work. Let $\Omega =\left( S_{min}, S_{max}\right)$ be an open subset of $\mathbb{R}^+$ and $\Omega_T \, = \, \Omega \times \left(0,T\right)$. Let $1\leq p \leq \infty$ and $k \, \in \, \mathbb{N}$. We define the following Sobolev spaces

\begin{align*}
W^{k}_{p}\left(\Omega\right) \, &= \, \left\lbrace u \, \in \, L^p\left(\Omega\right) \, \vert D^{\alpha}u \, \in \,  L^p\left(\Omega\right), 1\leq \left\lvert \alpha \right\rvert \leq k \right\rbrace,\\
W^{2k,k}_{p}\left(\Omega_T\right) \, &= \, \left\lbrace u \, \in \, L^p\left(\Omega_T\right) \, \vert D^{\alpha}\partial_t^{\beta}u \, \in \,  L^p\left(\Omega_T\right), 1\leq \left\lvert \alpha \right\rvert + 2\beta \leq 2k \right\rbrace,\\
\end{align*}

\noindent where $D^{\alpha}\partial_t^{\beta}u$ is the weak partial derivative of $u$. These spaces are actually Banach spaces when assigning the following norms

\begin{align*}
\left\lVert u \right\rVert_{W^{k}_{p}\left(\Omega\right)} &=\, \sum_{0\leq \left\lvert \alpha \right\rvert \leq k} \left\lVert D^{\alpha} u \right\rVert_{L^{p}\left(\Omega\right)},\\
\left\lVert u \right\rVert_{W^{2k,k}_{p}\left(\Omega_T\right)} &=\, \sum_{0\leq \left\lvert \alpha \right\rvert + 2\beta \leq 2k} \left\lVert D^{\alpha}\partial_t^{\beta}u \right\rVert_{L^{p}\left(\Omega\right)},
\end{align*}

\subsection{Existence}\label{existence}

\noindent We are going to look for convex solutions of the problem \eqref{eq7}. Problems of this kind refer to any derivative whose payoff correspond to a convex function as it could be an European option call. Hence, the modified volatility defined in \eqref{mod_vol} is changed to

\begin{align}
\hat{\sigma}^2= \sigma^2 \, \left(1 -  \sqrt{\frac{2}{\pi \, dt}} \,  \frac{C_S}{\sigma}  \right). \label{mod_vol_2}
\end{align}

\noindent Also, we apply the change of variables $x=\log\left(S\right)$ and $\tau=T-t$. We define the parabolic operator $\mathcal{L}$ and the nonlinear operator $\mathcal{N}$ as

\begin{align*}
\mathcal{L} \, V \, &= - \, \frac{\partial V}{\partial \tau} + \frac{1}{2} \, \hat{\sigma}^2 \, \frac{\partial^2 V}{\partial x^2} + \frac{\partial V}{\partial x} \, \left(q_S - \gamma_S  - \frac{1}{2}\hat{\sigma}^2\right) -  r \,V \\
\mathcal{N} \, V \, &=  V^{+} \left[s_F \, + \lambda_C \, \left(1-R_C \right) \right] + \,  V^{-} \, \left(\lambda_B - r \,C_B \right) \, \left(1-R_B \right)  + \sigma  \, \sqrt{\frac{2}{\pi \, dt}} \, C_C \, \left(1-R_C \right) \, \left\lvert \, \frac{\partial V}{\partial x}^{+} \right\rvert 
\end{align*}

\noindent such as the problem reads as

\begin{align}
\mathcal{L} \, \hat{V}\left(\tau,x\right) &= \mathcal{N} \, \hat{V}\left(\tau,x\right) \quad \text{in} \quad \Omega\times\left[0,T\right] \nonumber\\
\hat{V}\left(0,x\right) &= g\left(x\right) \quad \text{in} \quad \Omega \label{prob1} \\
\hat{V}\left(\tau,x\right) &=f\left(x\right) \quad \text{in} \quad \partial\Omega \times \left(0,T \right). \nonumber
\end{align}

\noindent where $g\left(x\right)$ is the initial condition (i.e. the payoff of the derivative) and  $f\left(x\right)$ is the boundary condition. For example, we define the conditions for an European call option as

\begin{align*}
g\left(x\right)=\left( \exp\left(x\right)-K \right)^+,
\end{align*}

\[
f\left(x\right)=  \left\{
													\begin{array}{ccc}
													0 & \hbox{if} & x \rightarrow 0\\
													\exp\left(x\right) & \hbox{if} & x \rightarrow \infty.\\
													\end{array}
													\right.
\]\\

\noindent In order to find a solution of problem \eqref{prob1}, we define an operator $T: \, C^{1,0}\left(\bar{\Omega}\right)\rightarrow C^{1,0}\left(\bar{\Omega}\right)$ such that $T\left(u\right)=v$, where $v \, \in \, W^{2,1}_{p}$ is the unique solution of the problem $\mathcal{L} v=\mathcal{N}u$. Our objective is to find a fixed point of the operator $T$ which at the same time will be the solution of problem \eqref{prob1}. We will set three conditions that the parameters of the model must fulfill to assess the existence of a convex solution.\\

\noindent The first condition is required to define a well-posed equation. As explained in \cite{leland1985option}, the modified volatility shown in equation \eqref{mod_vol_2} must be positive. This can be addressed by setting a lower bound for the volatility parameter as 

\begin{align}
\sigma > \sqrt{\frac{2}{\pi \, dt}} \, C_S \label{condition1}
\end{align}

\noindent The second one is a sufficient condition which is required to find a fixed point of the operator $T$. 
Let $c$ be positive constant depending only on the domain, to be defined below and assume
that the following inequality holds:

\begin{align}
c \, \left( \left[s_F \, + \lambda_C \, \left(1-R_C \right) \right]  \, + \, 2 \, \left(\lambda_B - r \,C_B \right) \, \left(1-R_B \right)  \, + \, \sigma  \, \sqrt{\frac{2}{\pi \, dt}} \, C_C \, \left(1-R_C \right) \right) < 1.
\end{align}

\noindent Given all the parameters of the model set, this assumption can be rewritten in terms of an upper bound for the volatility parameter

\begin{align}
\sigma \, < \, \frac{1-c \left( \left[s_F \, + \lambda_C \, \left(1-R_C \right) \right]  \, + \, 2 \, \left(\lambda_B - r \,C_B \right) \, \left(1-R_B \right) \right) }{c \, \sqrt{\frac{2}{\pi \, dt}} \, C_C \, \left(1-R_C \right)} \label{condition2}
\end{align}


\noindent The third and last condition shall be used to prove that the solution found is indeed convex. We shall assume that the stock growth rate under the risk neutral measure has to be bounded, more specifically:

\begin{align}
q_S-\gamma_S < M:=\max\left(r_1 + \sigma \, S_{max}  \, \sqrt{\frac{2}{\pi \, dt}} C_C \, \left(1-R_C \right), r_2 \right),
 \label{condition4}
\end{align}
\noindent where $r_1=r-\left[ s_F  + \lambda_C \, \left(1-R_C \right) \right]$ and $r_2=r-\left(\lambda_B - r C_B\right) \, \left(1-R_B \right)$.\\

\noindent The main theorem of the paper reads as follows.

\begin{theorem}\label{main-th}
Suppose that assumptions \eqref{condition1}, \eqref{condition2} and \eqref{condition4} hold, that both the initial and boundary conditions belong to the $W^{2,1}_{p}$ space and the initial condition is a convex function. Then, problem \eqref{prob1} admits at least one solution.
\end{theorem}

\noindent The main idea of the proof is to apply the Schauder fixed point theorem to the operator $T$ previously defined. To this end, we shall define a nonempty convex, closed and bounded $K\subset \Omega$ such that $T$ is a compact continuous mapping of $K$ into itself. \\

\noindent For a proof of Theorem \ref{main-th}, let us firstly recall Theorem 7.32 from \cite{lieberman1996second} 
which shows that the operator $T$ is well defined and provides a lower estimate for $\mathcal{L}u$. By adapting this result to our problem, we get the following lemma:

\begin{lemma}\label{lemma1}
Let $u \in C^{1,0}\left(\bar{\Omega}\right)$, $\mathcal{L}$ be as in Theorem 7.32 from \cite{lieberman1996second} and $f:= \mathcal{N}u$. Then there exists a unique solution of problem $\mathcal{L} v=f$ in $\Omega$,  $v=g$ in $\lbrace 0 \rbrace \times \Omega$ and $v=\varphi$ on $\partial\Omega \, \times \, \left(0,T \right)$. Moreover, there exists $C>0$ independent of $f$ such that $v$ satisfies the estimate

\begin{align}
\left\lVert v \right\rVert_{W^{2,1}_{p}} \leq C \, \left( \left\lVert \mathcal{L}v \right\rVert_{p} +\left\lVert \varphi \right\rVert_{W^{2,1}_{p}} + \left\lVert g \right\rVert_{W^{2,1}_{p}} \right).
\end{align}
\end{lemma}

\noindent The following lemma will be useful to address the continuity of the operator $T$.

\begin{lemma}\label{lemma2}
Let $p>N$ and $ u_n  \in W^{2,1}_{p}$ a bounded sequence such that $u_n \rightarrow u$ pointwise. Given $G\left(x\right)=\max\left(x,0\right)$ it follows that
\begin{align}
G\left(u_n\right) \rightarrow G\left(u \right) \quad \text{in} \quad L^{p} \nonumber\\
\frac{\partial G\left(u_n\right)}{\partial x} \rightarrow \frac{\partial G\left(u \right)}{\partial x} \quad \text{in} \quad L^{p}
\end{align}
\end{lemma}

\begin{proof1}
The proof of the first statement follows from noting that $\left\vert G^{\prime}\left(x\right) \right\vert \leq 1$ so then $\left\lVert G\left(u_n\right) - G\left(u\right) \right\rVert_p \leq \left\lVert u_n - u \right\rVert_p \rightarrow  0$. To address the second statement, we first note that 

\begin{align*}
\frac{\partial}{\partial x} \left(G \circ u \right)=\left(G^{\prime} \circ u \right) \frac{\partial u}{\partial x}.
\end{align*}

Then, we can rewrite

\begin{align*}
\left(G^{\prime} \circ u \right) \frac{\partial u}{\partial x} - \left(G^{\prime} \circ u_n \right) \frac{\partial u_n}{\partial x} = \left(G^{\prime} \circ u  - G^{\prime} \circ u_n \right) \frac{\partial u}{\partial x} + \left(G^{\prime} \circ u_n  \right) \left( \frac{\partial u}{\partial x}-\frac{\partial u_n}{\partial x}\right)
\end{align*}

For the first term, since $\left\vert G^{\prime}\left(u\right) - G^{\prime}\left(u_n\right) \right\vert \leq 1$ it follows that

\begin{align*}
\left\vert \left(G^{\prime}\left(u\right) - G^{\prime}\left(u_n\right) \right) \frac{\partial u}{\partial x} \right\vert^{p} \leq \left\vert \frac{\partial u}{\partial x} \right\vert^{p}
\end{align*}

Then, by dominated convergence theorem

\begin{align*}
\left\lVert \left(G^{\prime}\left(u\right) - G^{\prime}\left(u_n\right) \right) \frac{\partial u}{\partial x} \right\rVert_p^p = \int \left\vert \left(G^{\prime}\left(u\right) - G^{\prime}\left(u_n\right) \right) \frac{\partial u}{\partial x} \right\vert^{p} \rightarrow 0.
\end{align*}

To assess the second term, we consider the inclusion $W^{2,1}_{p} \hookrightarrow  W^{1,0}_{p}$ to obtain a convergent subsequence in $W^{1,0}_{p}$. Nonetheless, given that $\left\vert G^{\prime}\left(u_n\right) \right\vert \leq 1$, it follows that

\begin{align*}
\left\vert G^{\prime}\left(u_n\right) \left( \frac{\partial u}{\partial x}-\frac{\partial u_n}{\partial x} \right)  \right\vert^{p} \leq \left\vert \frac{\partial u}{\partial x}-\frac{\partial u_n}{\partial x} \right\vert^{p} \rightarrow 0.
\end{align*}

and hence

\begin{align*}
\left\lVert G^{\prime}\left(u_n\right) \left( \frac{\partial u}{\partial x}-\frac{\partial u_n}{\partial x} \right)  \right\rVert^{p} \leq \left\lVert \frac{\partial u}{\partial x}-\frac{\partial u_n}{\partial x} \right\rVert^{p} \rightarrow 0
\end{align*}\\

\end{proof1}

\noindent Given Lemma \ref{lemma1} and Lemma \ref{lemma2}, we can now address the proof of Theorem \ref{main-th}. 
In order to apply Schauder Fixed Point theorem to the operator $T$, we set $K= \overline{B_R\left(u_0\right)}$ for some $u_0$; then, we have to prove that $T$ is a compact continuous mapping of $K$ into itself. Within our proof we will set $u_0=0$.\\

\noindent Let us first verify the continuity of the operator $T$. Let $u_n,u \in W^{2,1}_{p}$ such that $u_n \rightarrow u$  pointwise. By recalling Lemma \ref{lemma1}, there exists a constant $C>0$ such that

\begin{align}
\left\lVert Tu_n - T_u \right\rVert_{W^{2,1}_{p}} &\leq C \,\left\lVert \mathcal{L}\,Tu_n - \mathcal{L}\,Tu \right\rVert_{p} \nonumber\\
&\leq C \,\left\lVert \mathcal{N}\,u_n - \mathcal{N}\,u \right\rVert_{p} \nonumber\\
&\leq C \, \left[s_F \, + \lambda_C \, \left(1-R_C \right) \right] \,\left\lVert u_n^{+} - u^{+} \right\rVert_{p} + \, C \, \left(\lambda_B - r \,C_B \right) \, \left(1-R_B \right) \, \left\lVert u_n^{-} - u^{-} \right\rVert_{p} \nonumber \\
&+ \, C \, \sigma  \, \sqrt{\frac{2}{\pi \, dt}} \, C_C \, \left(1-R_C \right) \, \left\lVert \frac{\partial u_n}{\partial x}^{+} - \frac{\partial u}{\partial x}^{+} \right\rVert_{p} \\
\end{align}

\noindent By applying Lemma \ref{lemma2}, we know that $\left\lVert u_n^+ - u^+ \right\rVert_{p} \rightarrow 0$ and $\left\lVert \frac{\partial u_n}{\partial x}^{+} - \frac{\partial u}{\partial x}^{+} \right\rVert_{p} \rightarrow 0$. The same lemma is valid by changing the function $G$ into $G\left(x\right)=\min\left(x,0\right)$ so that $\left\lVert u_n^- - u^- \right\rVert_{p} \rightarrow 0$.\\

\noindent The compactness of the operator is addressed by considering a bounded subset $S \subset C^{1,0}$. By definition, the subset $S$ belongs to $W^{2,1}_p$ and $T\left(S\right) \, \subset \, W^{2,1}_p$. Given that $p>1$, the inclusion $W^{2,1}_p \hookrightarrow C^{1,0}$ guarantees that $T\left(S\right) \subset C^{1,0}$ is compact.\\

\noindent Further, let $R$ be a positive number such that

\begin{align}
R \, > \, \frac{\left\lVert f \right\rVert_{W^{2,1}_{p}} + \left\lVert g \right\rVert_{W^{2,1}_{p}}}{1-c\left(\left[s_F \, + \lambda_C \, \left(1-R_C \right) \right]  \, + \, 2 \,  \left(\lambda_B - r \,C_B \right) \, \left(1-R_B \right)  \, + \, \sigma  \, \sqrt{\frac{2}{\pi \, dt}} \, C_C \, \left(1-R_C \right) \right)}
\end{align}

\noindent and $u$ such that $\left\lVert u  \right\rVert_{C^{1,0}} \leq R$. Then, there exists a constant $c_1>0$ given by the embedding $W^{2,1}_p \hookrightarrow C^{1,0}$ so

\begin{align}
\left\lVert Tu  \right\rVert_{C^{1,0}} \leq c_1 \left\lVert Tu \right\rVert_{W^{2,1}_p}.\label{twoterm}
\end{align}

\noindent Given the inequality presented in equation \eqref{twoterm}, we can use the result of Lemma \ref{lemma1}. Hence, there exists a constant $c_2>0$ such that 

\begin{align}
\left\lVert Tu  \right\rVert_{C^{1,0}} \leq c_1 \, c_2 \, \left(\left\lVert \mathcal{N}u  \right\rVert_p + \left\lVert f \right\rVert_{W^{2,1}_{p}} + \left\lVert g \right\rVert_{W^{2,1}_{p}} \right).\label{twoterm2}
\end{align}

\noindent By recalling \eqref{prob1}, the nonlinear term $\mathcal{N}$ is bounded by

\begin{align}
\left\lVert \mathcal{N}u  \right\rVert_{p} \, &\leq  \, \left[s_F \, + \lambda_C \, \left(1-R_C \right) \right] \,\left\lVert u^+ \right\rVert_{p} +  \, \left(\lambda_B - r \,C_B \right) \, \left(1-R_B \right) \, \left\lVert u^- \right\rVert_{p} \nonumber\\
&+  \, \sigma  \, \sqrt{\frac{2}{\pi \, dt}} \, C_C \, \left(1-R_C \right) \, \left\lVert \frac{\partial u}{\partial x}^{+} \right\rVert_{p} \nonumber \\
& < \left[s_F \, + \lambda_C \, \left(1-R_C \right) \right] \, R \, + \, \left(\lambda_B - r \,C_B \right) \, \left(1-R_B \right) \, 2R \, + \, \sigma  \, \sqrt{\frac{2}{\pi \, dt}} \, C_C \, \left(1-R_C \right) \, R \nonumber \\
&< k_1 \, R \label{firsterm}
\end{align}

\noindent using that $\left\lVert u  \right\rVert_{C^{1,0}} \leq R$ where $k_1=\left[s_F \, + \lambda_C \, \left(1-R_C \right) \right]  \, + \, \left(\lambda_B - r \,C_B \right) \, \left(1-R_B \right) \, 2 \, + \, \sigma  \, \sqrt{\frac{2}{\pi \, dt}} \, C_C \, \left(1-R_C \right)$.\\

%
%
%
%

\noindent By applying \eqref{firsterm} in \eqref{twoterm2}, we get that

\begin{align}
\left\lVert Tu  \right\rVert_{C^{1,0}} &\leq c_1 \, c_2 \, k_1  \, R + c_1 \, c_2 \, \left(\left\lVert f \right\rVert_{W^{2,1}_{p}} + \left\lVert g \right\rVert_{W^{2,1}_{p}}\right).\nonumber \\
&< \, R  \label{twoterm3} 
\end{align}

\noindent which follows from the assumption \eqref{condition2} by setting $c=c_1 \, c_2$ and the lower bound of $R$.\\

\noindent The last step of the proof is to show that the solution is indeed convex. To this end, we can analyze the similarity between equation \eqref{eq7} and a Black-Scholes equation with dividends and notice that given a convex initial condition, the solution would remain convex. We analyze separately when $\hat{V}$ is positive or negative.\\

\noindent When the solution is positive, equation \eqref{eq7} reduces to 

\begin{align*}
\frac{\partial \hat{V}}{\partial t} + \frac{1}{2} \, \hat{\sigma}^2 \, S^2 \, \frac{\partial^2 \hat{V}}{\partial S^2} + \frac{\partial \hat{V}}{\partial S} \, S \, \left(q_S - \gamma_S \right) -  r \, \hat{V} =& \, \left[ s_F  + \lambda_C \, \left(1-R_C \right) \right] \, \hat{V} \,  +  \sigma \, S  \, \sqrt{\frac{2}{\pi \, dt}} C_C \, \left(1-R_C \right) \, \left\lvert \frac{\partial \hat{V}}{\partial S} \right\rvert.
\end{align*} 

\noindent By rearranging terms and defining $r_1=r-\left[ s_F  + \lambda_C \, \left(1-R_C \right) \right]$ the equation above becomes

\begin{align}
\frac{\partial \hat{V}}{\partial t} + \frac{1}{2} \, \hat{\sigma}^2 \, S^2 \, \frac{\partial^2 \hat{V}}{\partial S^2} + \frac{\partial \hat{V}}{\partial S} \, S \, \left(q_S - \gamma_S - sgn\left(\frac{\partial \hat{V}}{\partial S} \right)\sigma \, S  \, \sqrt{\frac{2}{\pi \, dt}} C_C \, \left(1-R_C \right)  \right) -  r_1 \, \hat{V} = 0\label{positive_case}
\end{align} 

\noindent When the solution is negative, equation \eqref{eq7} reduces to

\begin{align*}
\frac{\partial \hat{V}}{\partial t} + \frac{1}{2} \, \hat{\sigma}^2 \, S^2 \, \frac{\partial^2 \hat{V}}{\partial S^2} + \frac{\partial \hat{V}}{\partial S} \, S \, \left(q_S - \gamma_S \right) -  r \, \hat{V} = \left(\lambda_B - r C_B\right) \, \left(1-R_B \right) \, \hat{V}.
\end{align*}

\noindent By rearranging terms and defining $r_2=r-\left(\lambda_B - r C_B\right) \, \left(1-R_B \right)$ the equation above becomes

\begin{align}
\frac{\partial \hat{V}}{\partial t} + \frac{1}{2} \, \hat{\sigma}^2 \, S^2 \, \frac{\partial^2 \hat{V}}{\partial S^2} + \frac{\partial \hat{V}}{\partial S} \, S \, \left(q_S - \gamma_S \right) -  r_2 \, \hat{V} = 0\label{negative_case}
\end{align}

\noindent Equation \eqref{positive_case} and \eqref{negative_case} can be thought as a Black-Scholes equation with dividend yield $\gamma_S$ and free-risk interest rate $r_1$ and $r_2$ respectively. Moreover, the condition stated in \eqref{condition4} can be used to derive an upper bound for $q_S-\gamma_S$. If $q_S-\gamma_S < M$, we see that the growth rate of the stock under the risk-free measure is lower than the free-risk interest rate. This dynamic is the one expected for a Black-Scholes model with dividends. Because the initial condition of the problem is indeed convex the solution $\hat{V}$ is also convex.

%
%
%
 

\section{Numerical Implementation}
\subsection{Numerical Framework}\label{Num_Frame}

In this section we develop a numerical framework to solve the problem defined in \eqref{prob1} by applying a forward Euler method. Hence, we recall the nonlinear problem 

\begin{align}
\mathcal{L} \, \hat{V}\left(\tau,x\right) &= \mathcal{N} \, \hat{V}\left(\tau,x\right) \quad \text{in} \quad \Omega\times\left[0,T\right] \nonumber\\
\hat{V}\left(0,x\right) &= g\left(x\right) \quad \text{in} \quad \Omega \label{prob2} \\
\hat{V}\left(\tau,x\right) &=f\left(x\right) \quad \text{in} \quad \partial\Omega \times \left(0,T \right). \nonumber
\end{align}

\noindent with $\mathcal{L}$ and $\mathcal{N}$ defined in \eqref{prob1}. For numerical convenience, we approximate the original smooth domain by a discrete one $\hat{\Omega}_T\subset\left[ a,b \right] \times \left[ 0,T \right]$, setting $a$ and $b$ in order to cover a set of feasible logarithmic stock prices. The step of the temporal variable is uniformly set as $\Delta \tau=T/T_x$ being $T_x$ the number of grid points in the $\tau$- direction. For the spatial variable, we decide to apply a non-uniform grid where the spacing is fine near the strike and coarse away from the strike. In \cite{tavella2000pricing}, the following grid is proposed

\begin{align}
x_i=x^* \, + \alpha \, \sinh\left(c_2 \frac{i}{N} \, + \, c_1 \left(1- \frac{i}{N} \right) \right) 
\end{align}

\noindent where

\begin{align*}
c_1 &= \sinh^{-1} \left(\frac{x^{-}-x^*}{\alpha} \right)\\
c_2 &= \sinh^{-1} \left(\frac{x^{+}-x^*}{\alpha} \right).
\end{align*}

\noindent This is a transformation that maps the interval $[0,1]$ into $[x^-,x^+]$ by concentrating the points near $x^*$. The value of $\alpha$ sets how non-uniform the grid will be and $N$ to be the amount of points within the grid. In our problem we set $x^*= K$ and $[x^-,x^+]$ accordingly to cover all the possible logarithmic prices. Hence, we define the solution to the $m$-temporal step as $\hat{V}^{m}_{i}=\hat{V}\left(x_i,m\Delta\tau\right)$ where $1\leq i \leq N$ and $1\leq m \leq T_x$. We also define $\hat{U}=\max\left(\hat{V},0\right)$ for numerical notation convenience.\\

\noindent To derive the expression of the numerical framework we follow \cite{bodeau2000non} and \cite{foulon2010adi} in which this grid had been applied. By following the same steps, we obtain that the discretization of the first and second spatial derivatives are given by

\begin{align*}
\frac{\partial \hat{V}}{\partial x}\, &= \,  \frac{\hat{V}^{m}_{i+1} - \hat{V}^{m}_{i}}{x_{i+1}-x_i},\\
\frac{\partial^2 \hat{V}}{\partial x^2}\, &= \,  h_i^+ \, \frac{\hat{V}^{m}_{i+1} - \hat{V}^{m}_{i}}{x_{i+1}-x_i} \, - \, h_i^- \, \frac{\hat{V}^{m}_{i} - \hat{V}^{m}_{i-1}}{x_{i}-x_{i-1}}
\end{align*}

\noindent where $h_i=x_i - x_{i-1}$ and

\begin{align*}
h_i^+ \, &= \, \frac{2}{h_{i+1} \, \left(h_{i+1} + h_i\right)},\\
h_i^- \, &= \, \frac{2}{h_{i} \, \left(h_{i+1} + h_i\right)}.
\end{align*}

\noindent Given that the temporal step is set uniformly, the finite difference framework is defined below

\begin{align}
\mathcal{L} \, \hat{V} \, =& \,  - \left(\frac{\hat{V}^{m+1}_{i} - \hat{V}^m_i}{\Delta \tau} \right) \, + \, \frac{1}{2} \, \hat{\sigma}^2 \, \left[h_i^+ \left(\hat{V}^{m}_{i+1} - \hat{V}^m_i \right) \, - \, h_i^- \, \left(\hat{V}^{m}_i - \hat{V}^{m}_{i-1} \right) \right] \, + \, \frac{\hat{V}^m_{i+1} - \hat{V}^m_i}{x_{i+1}-x_i} \nonumber \\
 &\left(q_S - \gamma_S-\frac{1}{2} \, \hat{\sigma}^2 \right) - \, r \, \hat{V}^m_i.\\
\mathcal{N} \, \hat{V} \, =& \,  \max\left(\hat{V}^m_i,0 \right) \left[s_F \, + \lambda_C \, \left(1-R_C \right) \right] + \, \min\left(\hat{V}^m_i,0 \right) \, \left(\lambda_B - r \,C_B \right) \, \left(1-R_B \right) \nonumber \\
 &+ \sigma  \, \sqrt{\frac{2}{\pi \, dt}} \, C_C \, \left(1-R_C \right) \, \left\lvert \, \frac{\hat{U}^m_{i+1} - \hat{U}^m_i}{x_{i+1}-x_i} \right\rvert. 
\end{align}

\noindent By rearranging and combining terms we obtain the following iterative process

\begin{align}
\hat{V}^{m+1}_i \, &= \, \hat{V}^m_i \, \left(1 \, - \, \frac{\hat{\sigma}^2 \, \Delta \tau}{2} \left(h_i^+ \, + \, h_i^-  \right) \, - \, \frac{\Delta \tau}{x_{i+1} - x_i} \left(q_S-\gamma_S-\frac{\hat{\sigma}^2}{2} \right) - r \, \Delta \tau  \right) \, + \, \hat{V}^m_{i-1} \, \left(\frac{\hat{\sigma}^2 \, \Delta \tau}{2} h_i^- \right) \nonumber \\
&+ \, \hat{V}^m_{i+1} \, \left(\frac{\hat{\sigma}^2 \, \Delta \tau}{2} h_i^+ \, + \, \frac{\Delta \tau}{x_{i+1} - x_i} \left(q_S-\gamma_S-\frac{\hat{\sigma}^2}{2} \right) \right) - \Delta \tau \, \max\left(\hat{V}^m_i,0 \right) \left[s_F \, + \lambda_C \, \left(1-R_C \right) \right] \nonumber \\
&- \Delta \tau \, \min\left(\hat{V}^m_i,0 \right) \, \left(\lambda_B - r \,C_B \right) \, \left(1-R_B \right) - \, \Delta \tau \, \sigma  \, \sqrt{\frac{2}{\pi \, dt}} \, C_C \, \left(1-R_C \right) \, \left\lvert \, \frac{\hat{U}^m_{i+1} - \hat{U}^m_i}{x_{i+1}-x_i} \right\rvert. 
\end{align}

\noindent If we let $\hat{V}^{0}_i \, = \, g\left(x_i\right)$, this framework can be used to find the solution of the problem \eqref{prob2} at each time step $m$. Further, results regarding the convergence, stability and consistency of the method adapted to non-uniform grids on Black-Scholes problems can be found in \cite{bodeau2000non} and \cite{in2009stability}.

\subsection{Numerical Analysis}

\noindent In this section we analyze the behavior of the option price for an European call under different scenarios. We perform a sensitivity analysis on the volatility, free-risk interest rate, transaction costs, recovery rates and hazard rates by stressing its values. Nonetheless, we compare our results with the ones obtained by the original model proposed in \cite{burgard2011partial} and calculate how the transaction costs impact the final CVA value. To further analyze the behavior of the option price, we calculate its derivatives with respect to certain parameters. This derivatives are known as Greeks and consist of Delta (derivative with respect to the option price), Gamma (second derivative with respect to the option price), Vega (derivative with respect to the volatility) and Rho (derivative with respect to the interest rate).\\

\noindent For notation purposes we recall $BK$ to the original model and $BK_{TC}$ the model with transaction costs. Also, for each scenario, the parameters set for both models are defined in the caption of each figure and results are obtained at time $\tau=T$. Within each figure, two types of vertical lines are included. The grey-shaded lines correspond to the non-uniform $x_i$ grid defined in Section \ref{Num_Frame} and the black dashed-line represents the strike value.\\

\subsubsection{Delta and Gamma}

\begin{figure}[h]
\centering
\begin{subfigure}{.5\textwidth}
  \centering
  \includegraphics[scale=0.5]{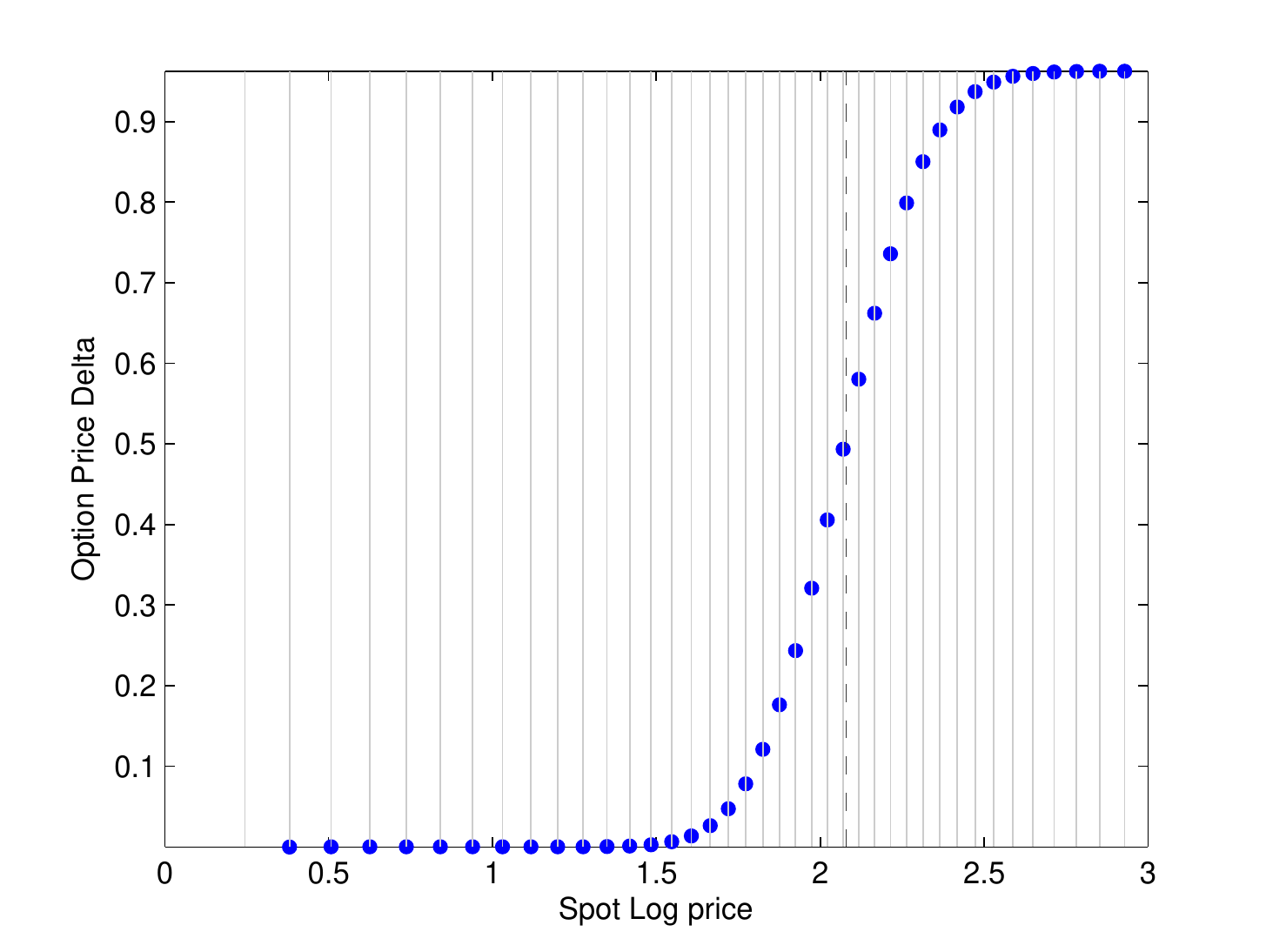}
  \caption{Option Delta}
  \label{fig:sub1}
\end{subfigure}%
\begin{subfigure}{.5\textwidth}
  \centering
  \includegraphics[scale=0.5]{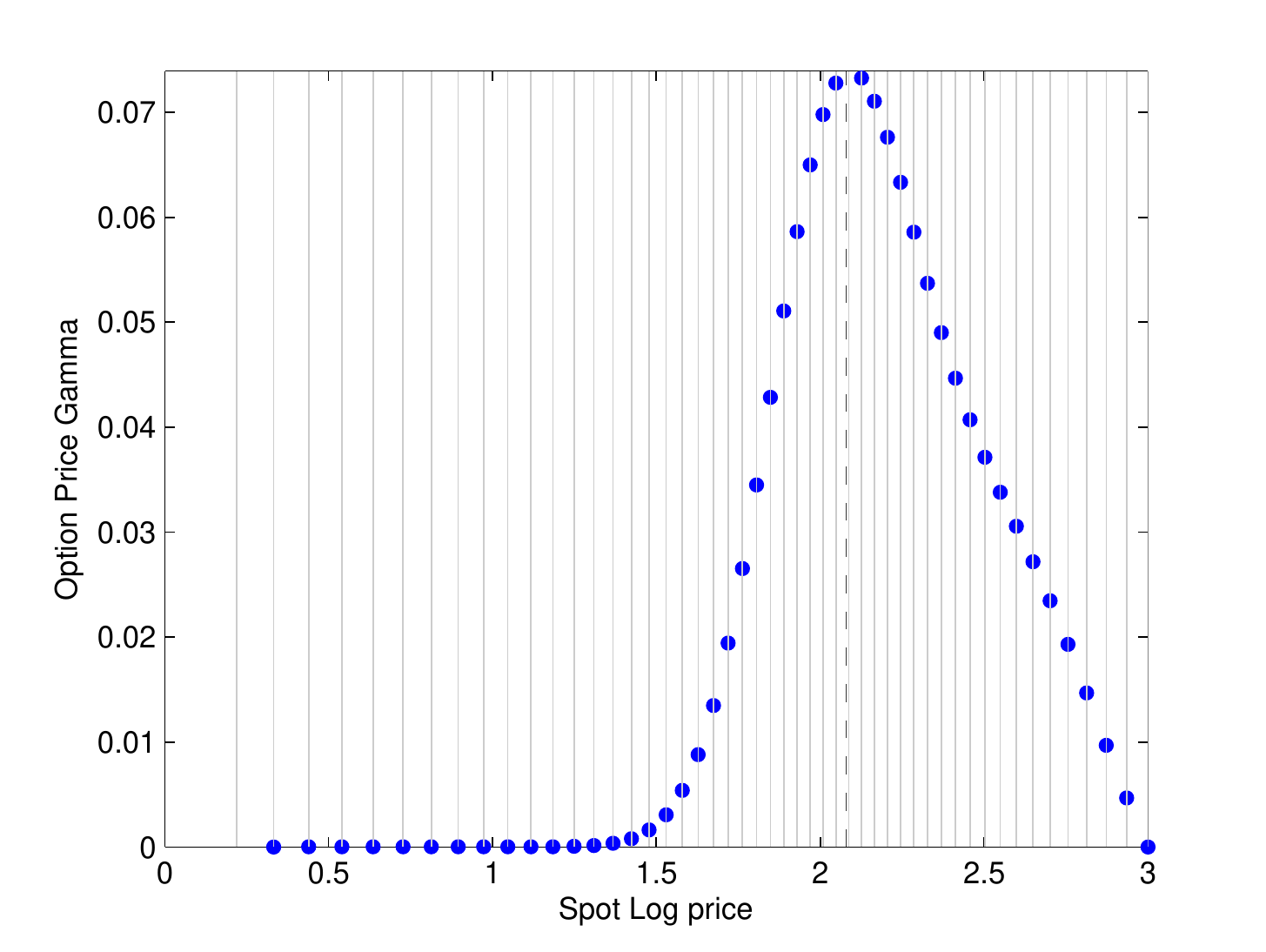}
  \caption{Option Gamma}
  \label{fig:sub2}
\end{subfigure}
\caption{$C_S =  0.002$, $r=0.05$, $q_S=0.05$, $\gamma_S=0.03$, $\sigma=0.1$, $S_f=0$, $\lambda_B=0.05$, $\lambda_C = 0.01$, $R_B=0.4$, $R_C = 0.4$, $C_B=0.001$, $C_C=0.001$, $dt=1/261$, $\Delta \tau=1/261$, $K=8$.}
\label{fig:deltagamma}
\end{figure}

\noindent Figure \eqref{fig:deltagamma} presents the two derivatives with respect to the option price, which are Delta and Gamma. Delta shows a similar behavior to an European call. It is known that for that vanilla option, Delta's formula correspond to a normal cumulative function. When including CVA and transaction costs, it can be seen that when the option is deep out-of-the-money, Delta is near zero which implies that the portfolio defined in Equation \eqref{rep_port} needs no shares of $S$ to hedge the option. As the option gets at-the-money, Delta grows approximately up to $0.5$. The option is more sensitive to changes in the spot price so then almost $50 \%$ of the hedging portfolio has to be covered with shares of $S$. This trend continues to converge to a Delta equal to $1$ when the option gets deeper in-the-money. At this point, the option price changes at the same rate with respect to the spot price and hedging portfolio has to only be long shares of $S$ to cover its hedging purpose.\\

\noindent Since Gamma represents the second derivative of the option price with respect to the spot price, its maximum is actually reached when the option is at-the-money and diminishes when the option go either in-the-money or out-of-the-money. This behavior is again similar to the one seen on a vanilla European call and shows to us how sensitive is Delta to movements in the spot price.\\

\subsubsection{Volatility}\label{subsub:vol}

\begin{figure}[hbtp]
\centering
\begin{subfigure}{.5\textwidth}
  \centering
  \includegraphics[scale=0.5]{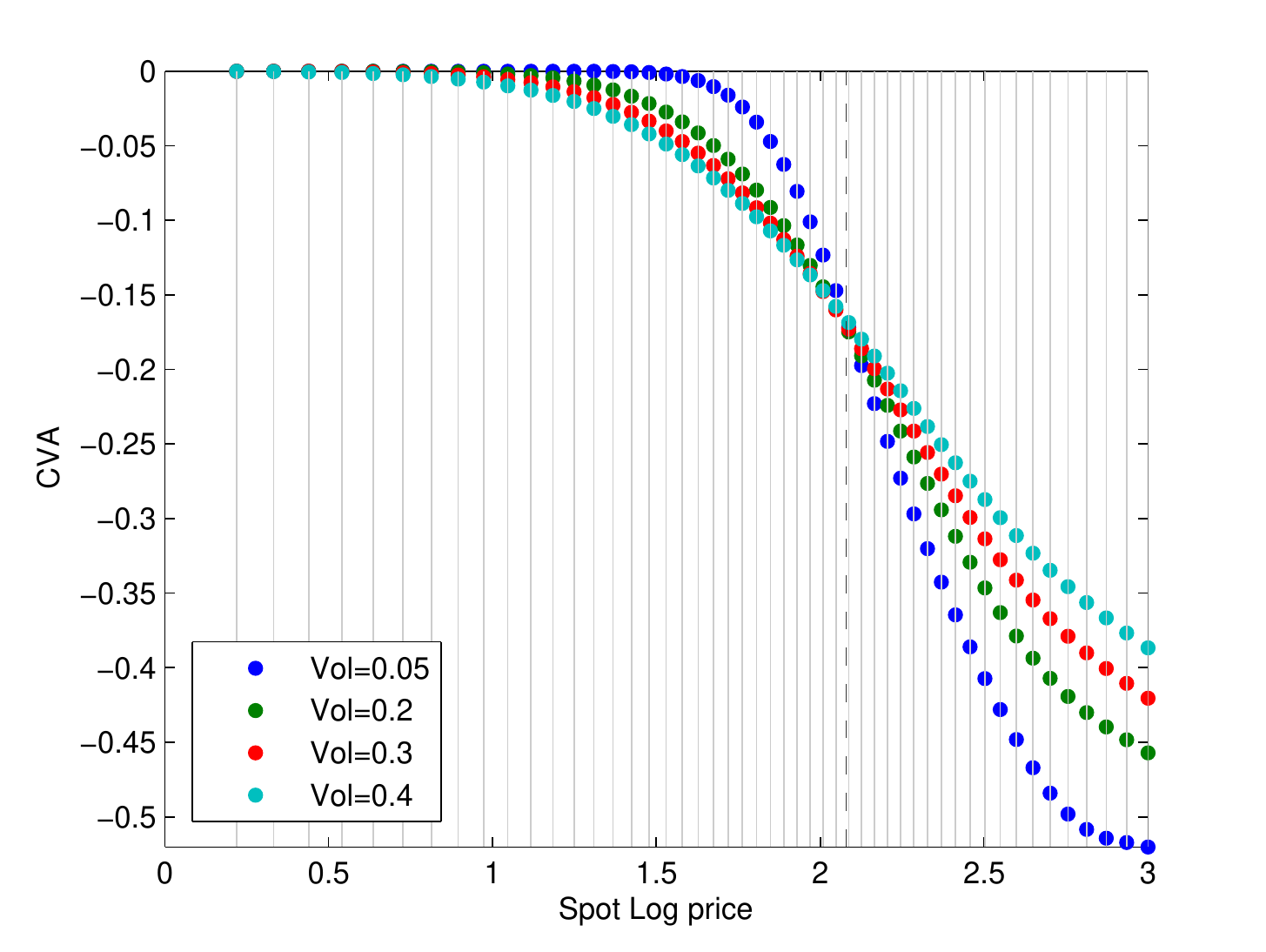}
  \caption{$BK_{TC}$ CVA for different volatilities}
  \label{fig:sub3}
\end{subfigure}%
\begin{subfigure}{.5\textwidth}
  \centering
  \includegraphics[scale=0.5]{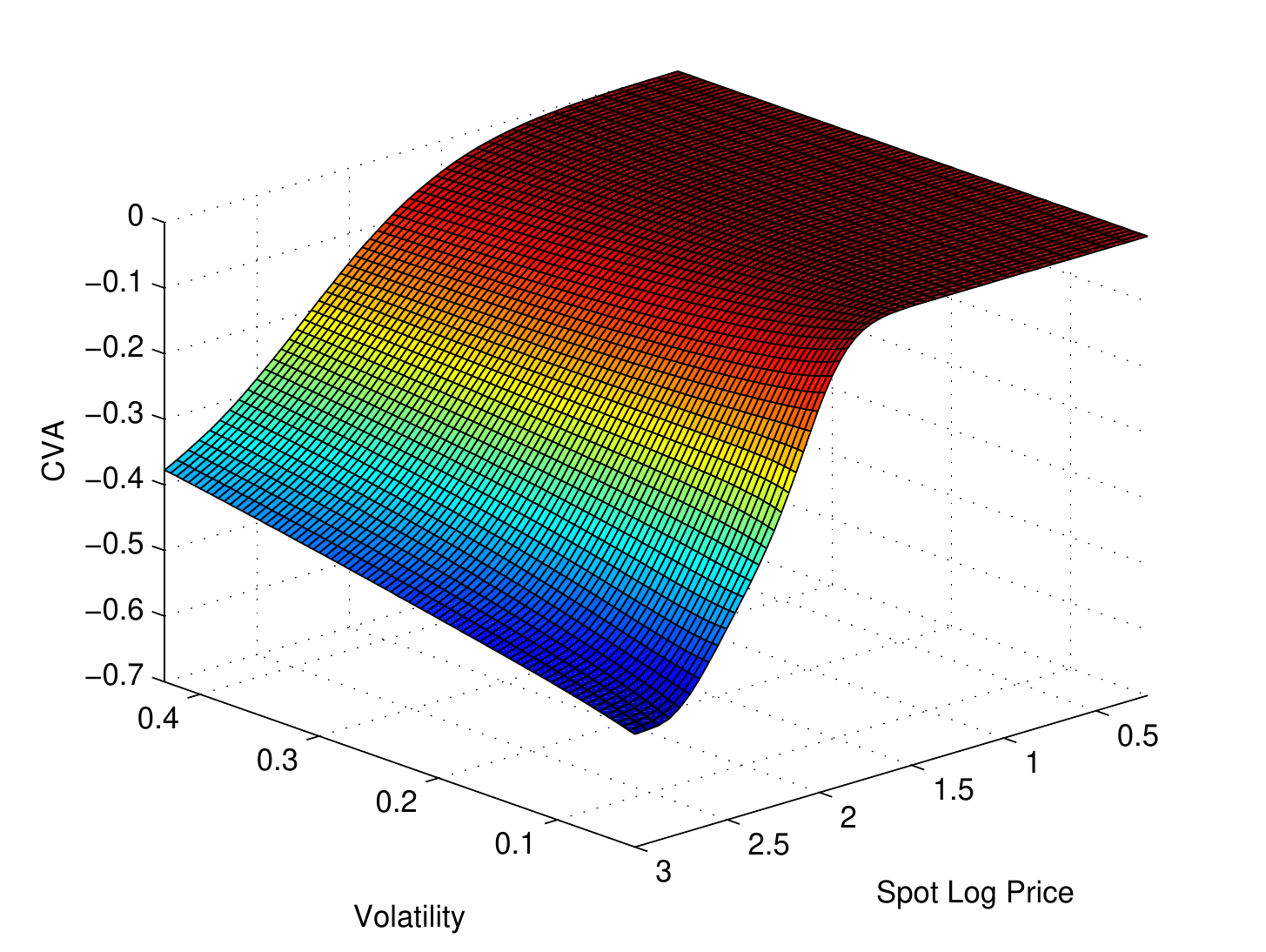}
  \caption{$BK_{TC}$ CVA by Spot Price and Volatility}
  \label{fig:sub5}
\end{subfigure}
\caption{$C_S =  0.002$, $q_S=0.05$, $\gamma_S=0.03$, $r=0.05$, $S_f=0$, $\lambda_C = 0.01$, $R_C = 0.4$, $\lambda_B=0.05$, $R_B=0.4$, $C_B=0.001$, $C_C=0.001$, $dt=1/261$, $\Delta \tau=1/261$, $K=8$.}
\label{fig:vol}
\end{figure}

\noindent Figure \eqref{fig:sub3} presents the sensitivity of the CVA to changes in the volatility parameter. The figure shows that the strike price serves as threshold where the behavior of the CVA changes. When the option is out-of-the money ($S<K$), higher volatility produces higher CVA (more negative). However, when the option is in-the-money ($S>K$), the convexity changes leading to higher CVA as the volatility decreases. Figure \eqref{fig:sub5} expands these results over the entire set of possible volatilities.\\

\begin{figure}[hbtp]
\centering
\begin{subfigure}{.5\textwidth}
  \centering
  \includegraphics[scale=0.5]{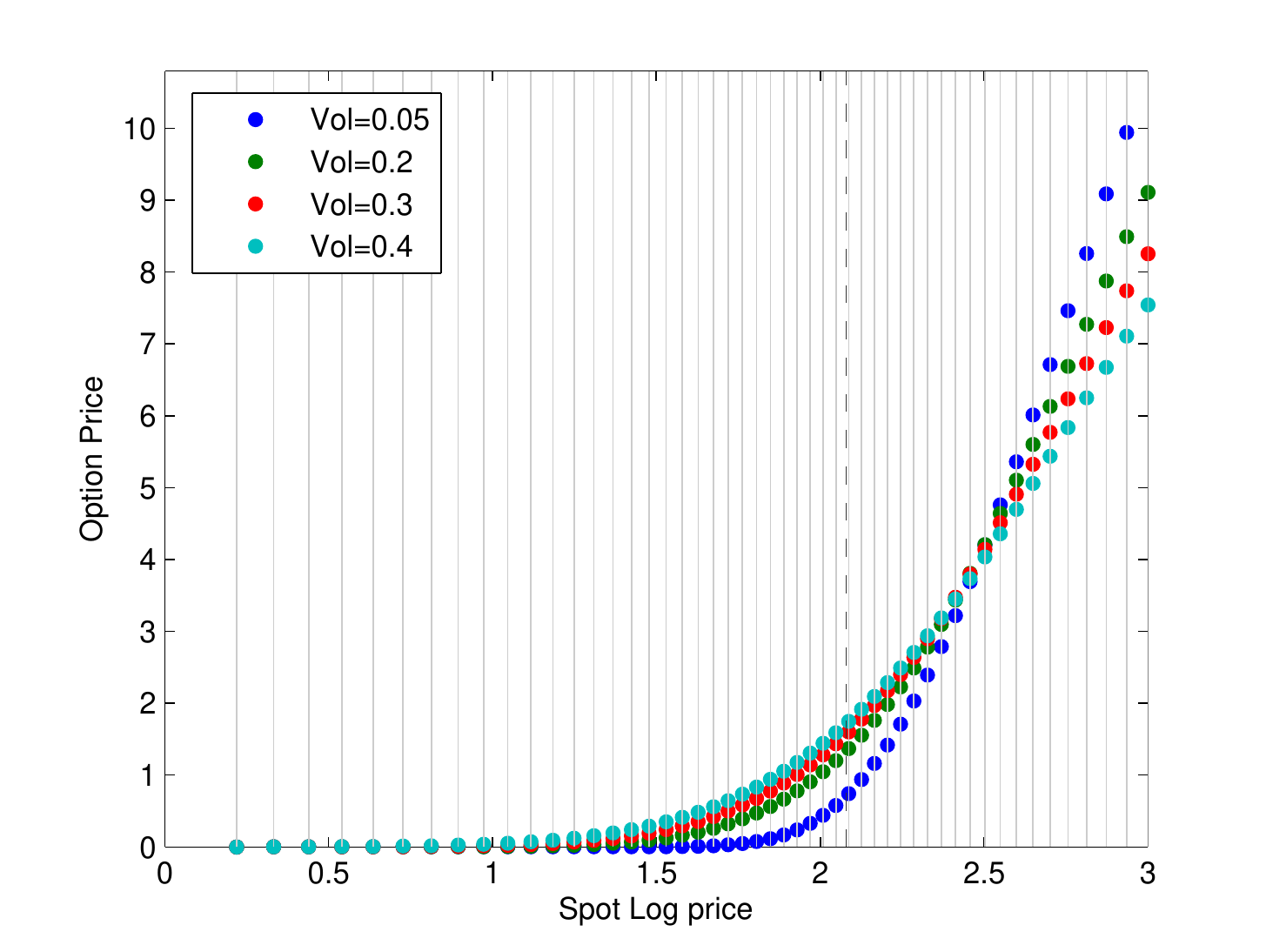}
  \caption{$BK_{TC}$ Option Price for different volatilities}
  \label{fig:sub4}
\end{subfigure}%
\begin{subfigure}{.5\textwidth}
	\centering
	\includegraphics[scale=0.5]{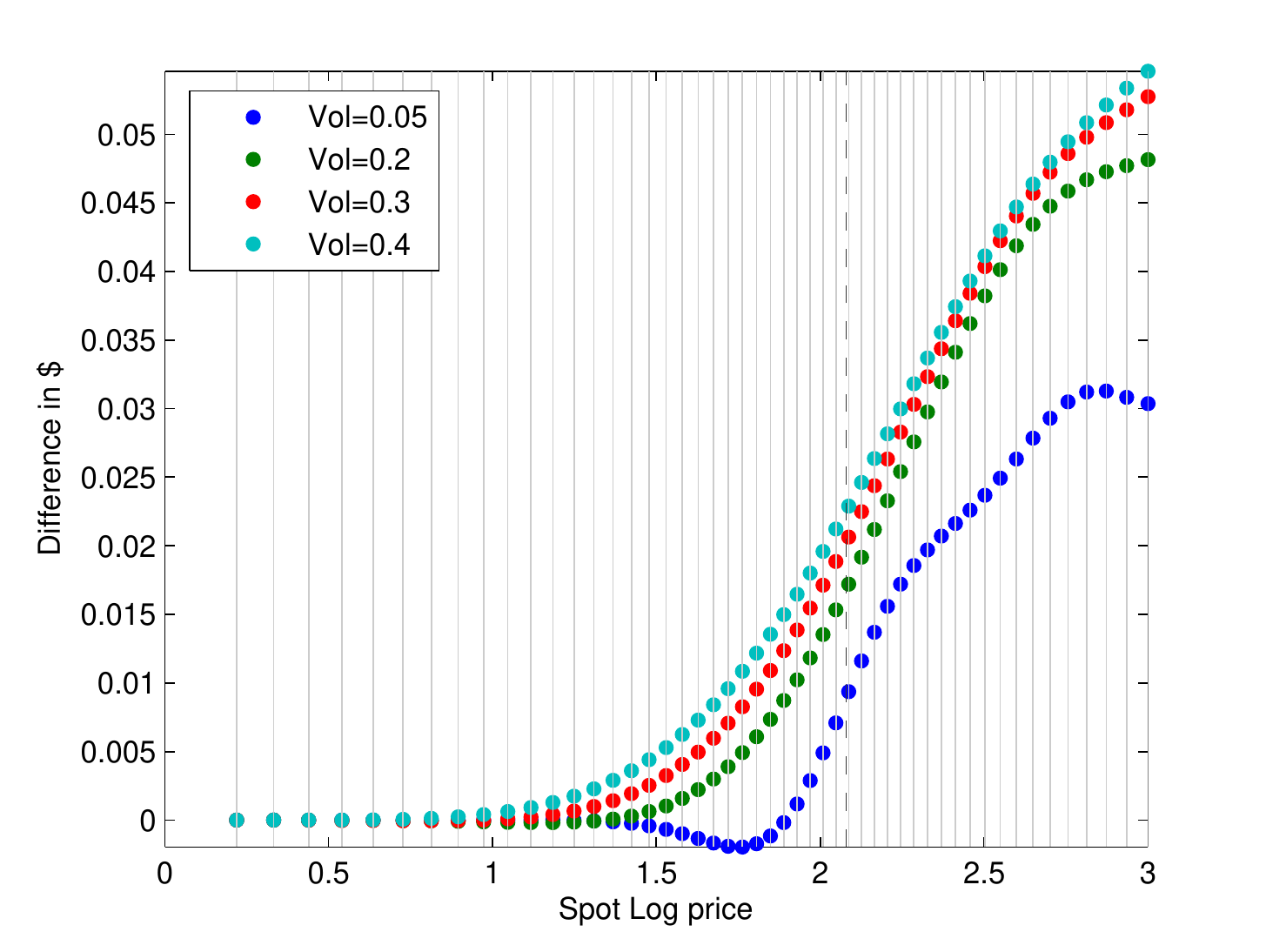}
	\caption{Difference between BK and $BK_{TC}$}
	\label{fig:sub6}
\end{subfigure}
\caption{$C_S =  0.002$, $q_S=0.05$, $\gamma_S=0.03$, $r=0.05$, $S_f=0$, $\lambda_C = 0.01$, $R_C = 0.4$, $\lambda_B=0.05$, $R_B=0.4$, $C_B=0.001$, $C_C=0.001$, $dt=1/261$, $\Delta \tau=1/261$, $K=8$.}
\label{fig:vol2}
\end{figure}

\begin{figure}[hbtp]
\centering
\begin{subfigure}{.5\textwidth}
  \centering
  \includegraphics[scale=0.5]{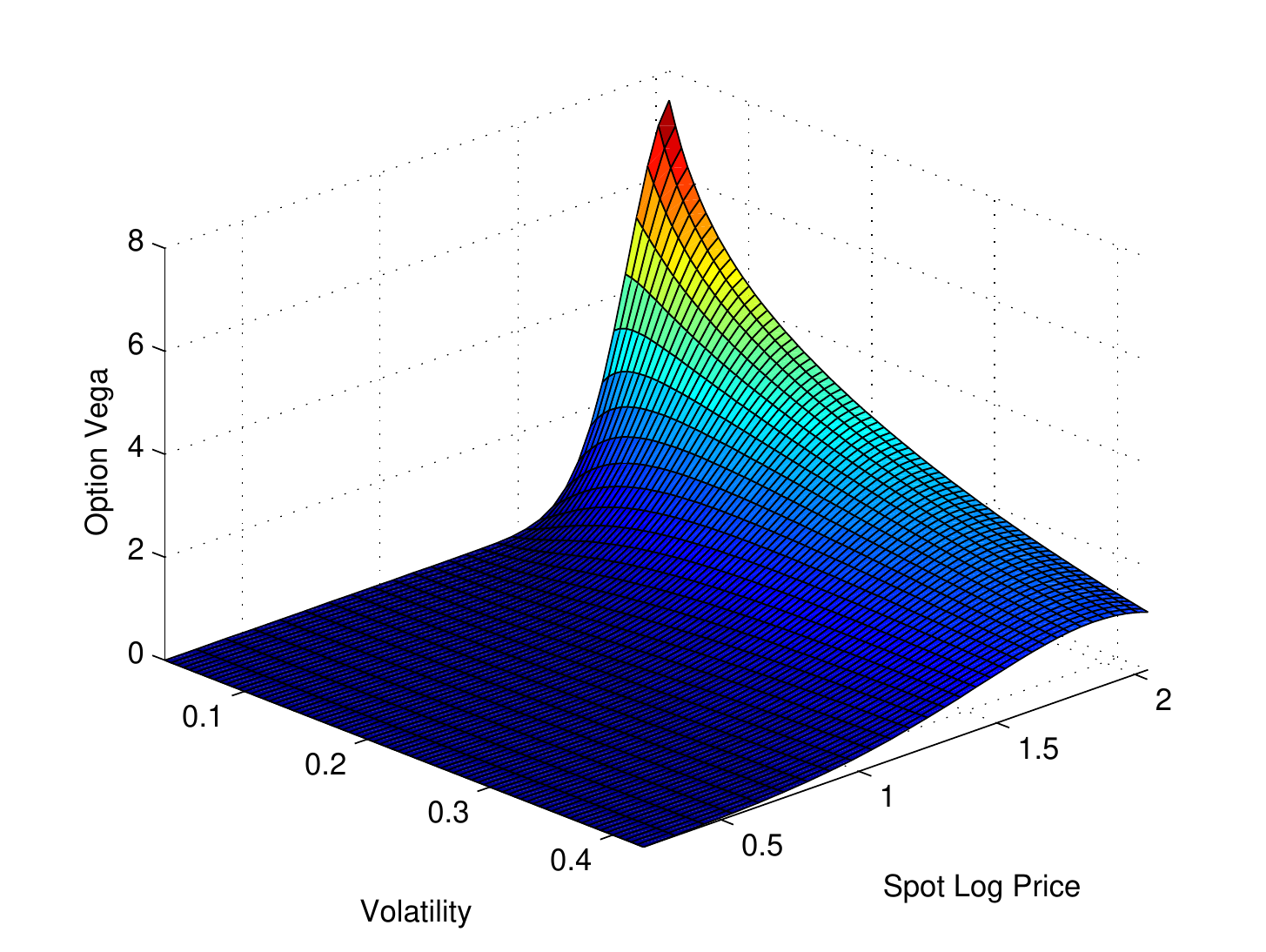}
  \caption{Option Vega out-of-the-money}
  \label{fig:sub7}
\end{subfigure}%
\begin{subfigure}{.5\textwidth}
  \centering
  \includegraphics[scale=0.5]{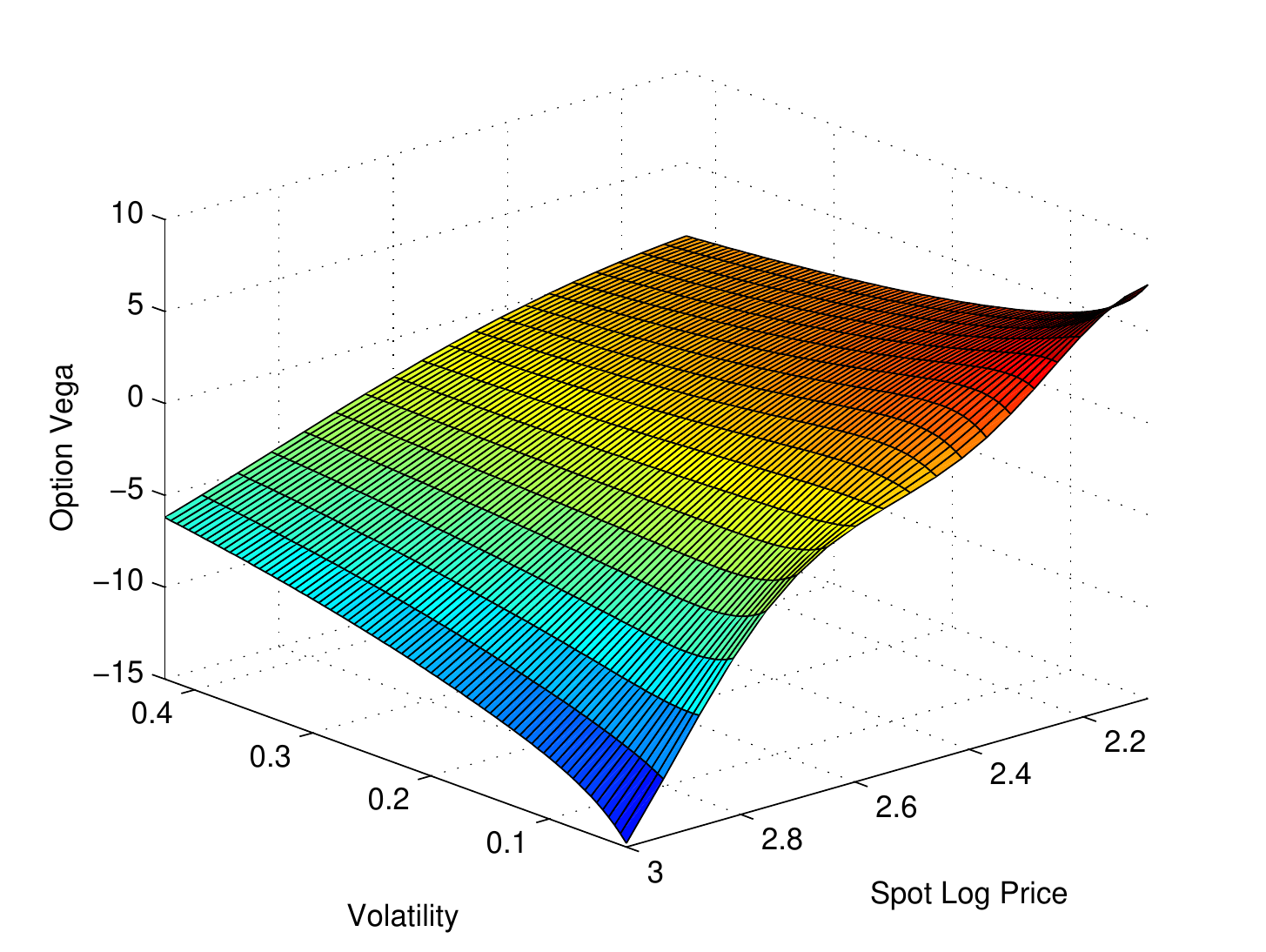}
  \caption{Option Vega in-the-money}
  \label{fig:sub8}
\end{subfigure}
\caption{$C_S =  0.002$, $r=0.05$, $q_S=0.05$, $\gamma_S=0.03$, $\sigma=0.1$, $S_f=0$, $\lambda_B=0.05$, $\lambda_C = 0.01$, $R_B=0.4$, $R_C = 0.4$, $C_B=0.001$, $C_C=0.001$, $dt=1/261$, $\Delta \tau=1/261$, $K=8$.}
\label{fig:vega}
\end{figure}

\noindent In Figure \eqref{fig:sub4}, the sensitivity of the option price with respect to different volatility parameters is presented. Under the usual Black-Scholes framework, it is expected to get higher option prices as volatility increases. This pattern is confirmed up to a certain spot price. Under our framework, as the option gets deeper in-the-money, Delta (Figure \eqref{fig:sub1}), which represents the amount of shares to buy in the replicant strategy, tends to $1$ and the impact of the transaction costs increase by generating a decrease in the option price. This behavior can be confirmed by assessing the first derivative of the option price with respect to the volatility (usually known as Vega). In Figure \eqref{fig:vega}, Vega is split with respect to the moneyness of the option. Figure \eqref{fig:sub7} shows that, when the option is out-of-the-money, Vega is positive as it is under the Black-Scholes model. Further, Figure \eqref{fig:sub8} demonstrate that not only Vega becomes negative as the option gets in-the-money but also that its sign changes in the same spot price as seen in Figure \eqref{fig:sub4}. Hence, if we consider the impact of the volatility not only in the parabolic side of the PDE but also in the nonlinear term , it is expected to find these relationship between the volatility parameter and the option price.\\

\subsubsection{Interest Rate}

\begin{figure}[h]
\centering
\begin{subfigure}{.5\textwidth}
  \centering
  \includegraphics[scale=0.5]{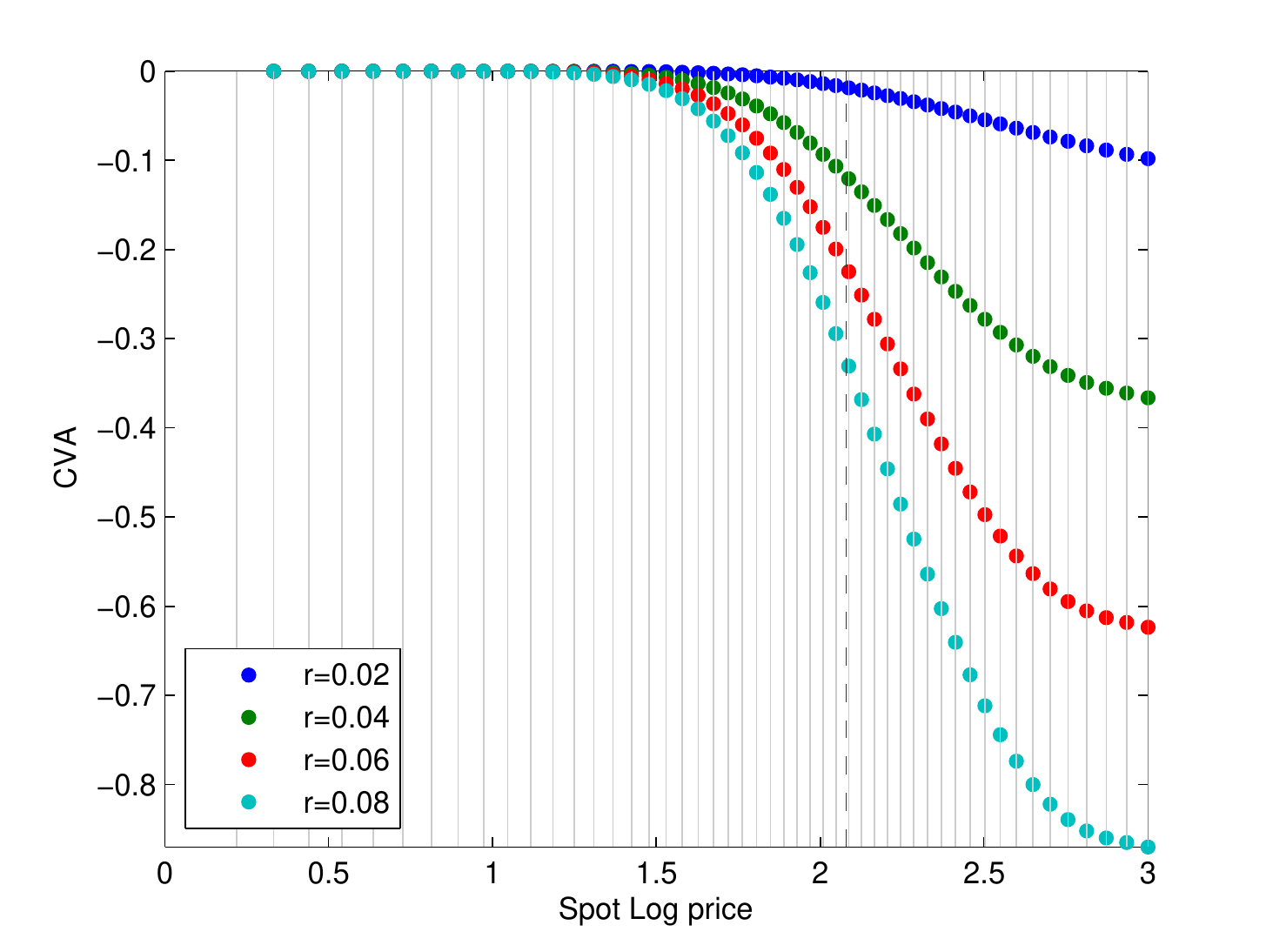}
  \caption{$BK_{TC}$ CVA for different interest rates.}
  \label{fig:sub9}
\end{subfigure}%
\begin{subfigure}{.5\textwidth}
  \centering
  \includegraphics[scale=0.5]{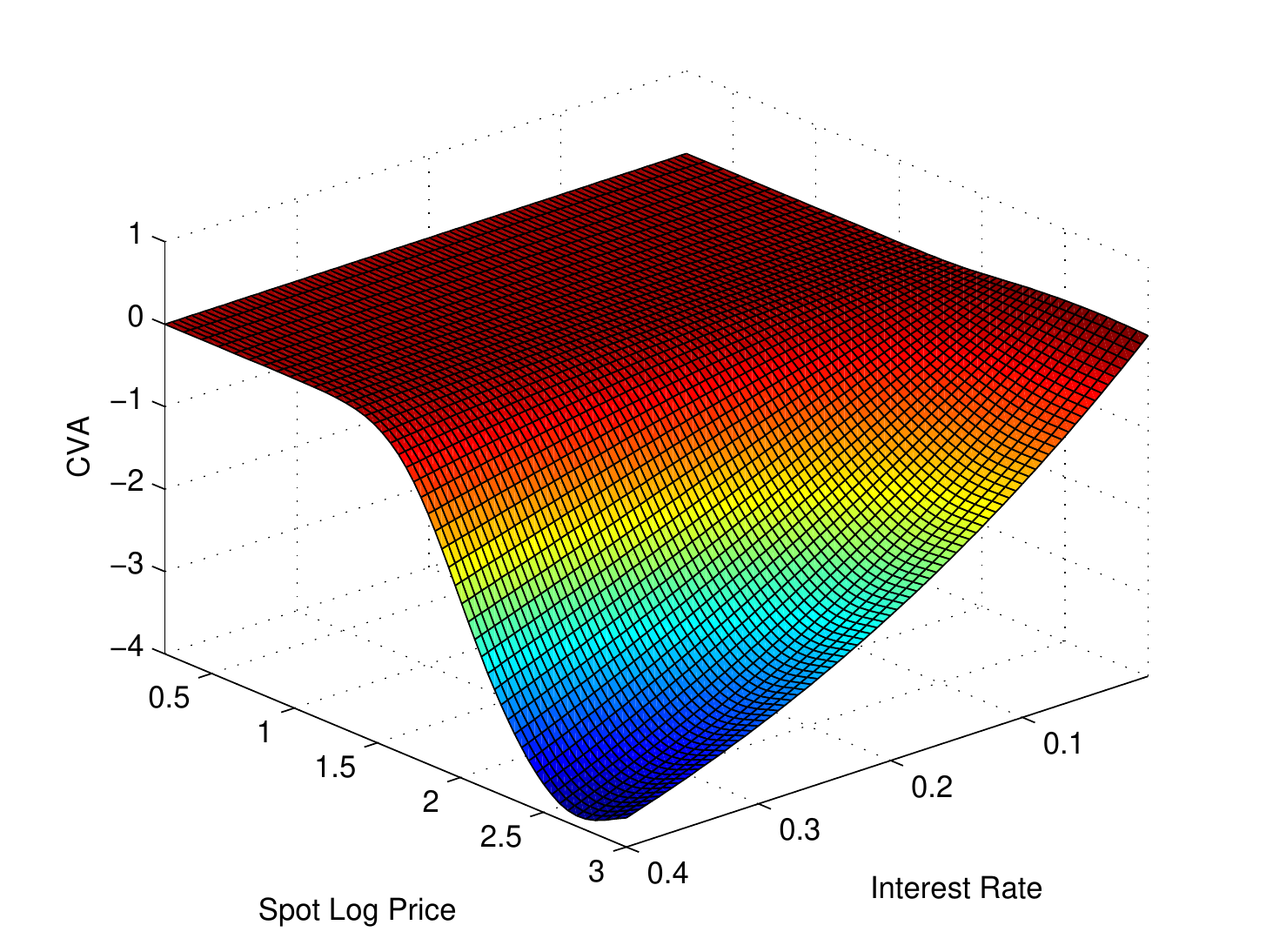}
  \caption{$BK_{TC}$ CVA by Spot price and interest rate.}
  \label{fig:sub10}
\end{subfigure}
\caption{$C_S =  0.002$, $q_S=0.05$, $\gamma_S=0.03$, $\sigma=0.1$, $S_f=0$, $\lambda_C = 0.01$, $R_C = 0.4$, $\lambda_B=0.05$, $R_B=0.4$, $C_B=0.001$, $C_C=0.001$, $dt=1/261$, $\Delta \tau=1/261$, $K=8$.}
\label{fig:r}
\end{figure}

\noindent Figure \eqref{fig:sub9} presents the sensitivity of the CVA to changes in the interest rate. Figure \eqref{fig:sub10} expand these results to the entire interval of Spot Log prices. Both figures show that the CVA decreases as the interest rate increases and the its size is larger when the option is deep in-the-money.\\

\begin{figure}[h]
\centering
\begin{subfigure}{.5\textwidth}
  \centering
  \includegraphics[scale=0.5]{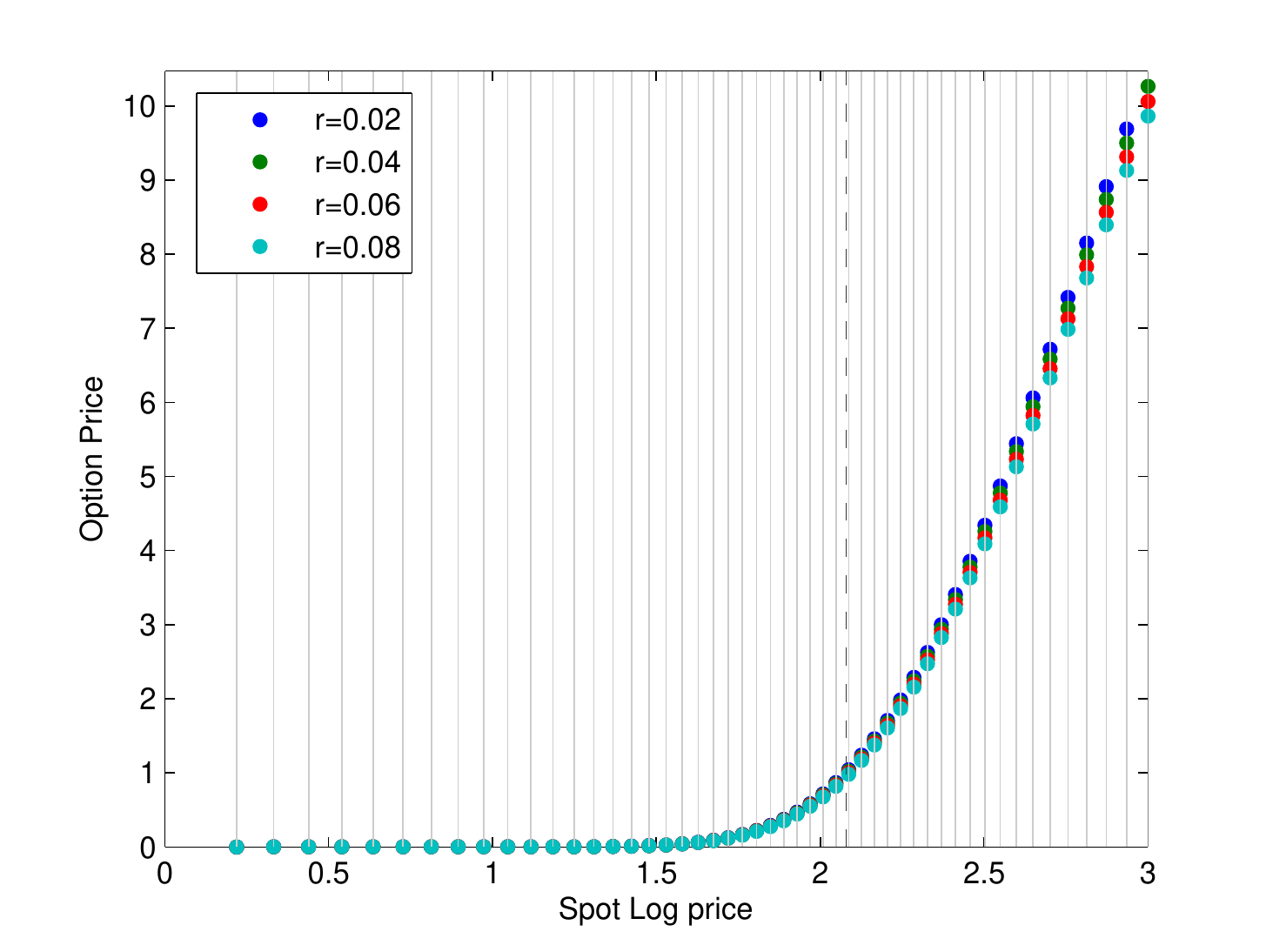}
  \caption{$BK_{TC}$ Option Price for different interest rates}
  \label{fig:sub11}
\end{subfigure}%
\begin{subfigure}{.5\textwidth}
  \centering
  \includegraphics[scale=0.5]{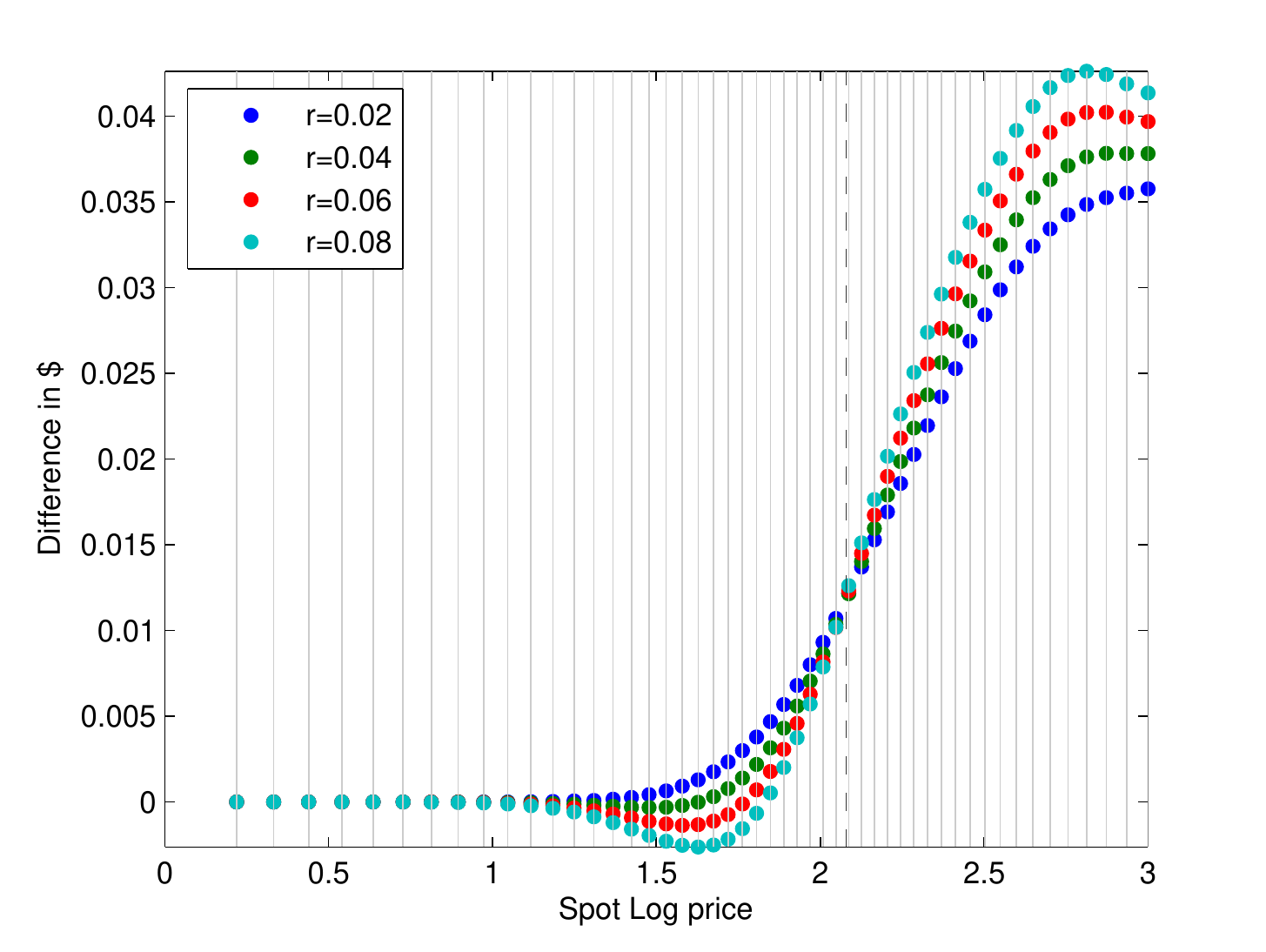}
  \caption{Difference between BK and $BK_{TC}$}
  \label{fig:sub12}
\end{subfigure}
\caption{$C_S =  0.002$, $q_S=0.05$, $\gamma_S=0.03$, $\sigma=0.1$, $S_f=0$, $\lambda_C = 0.01$, $R_C = 0.4$, $\lambda_B=0.05$, $R_B=0.4$, $C_B=0.001$, $C_C=0.001$, $dt=1/261$, $\Delta \tau=1/261$, $K=8$.}
\label{fig:r2}
\end{figure}

\begin{figure}[h]
  \centering
  \includegraphics[scale=0.5]{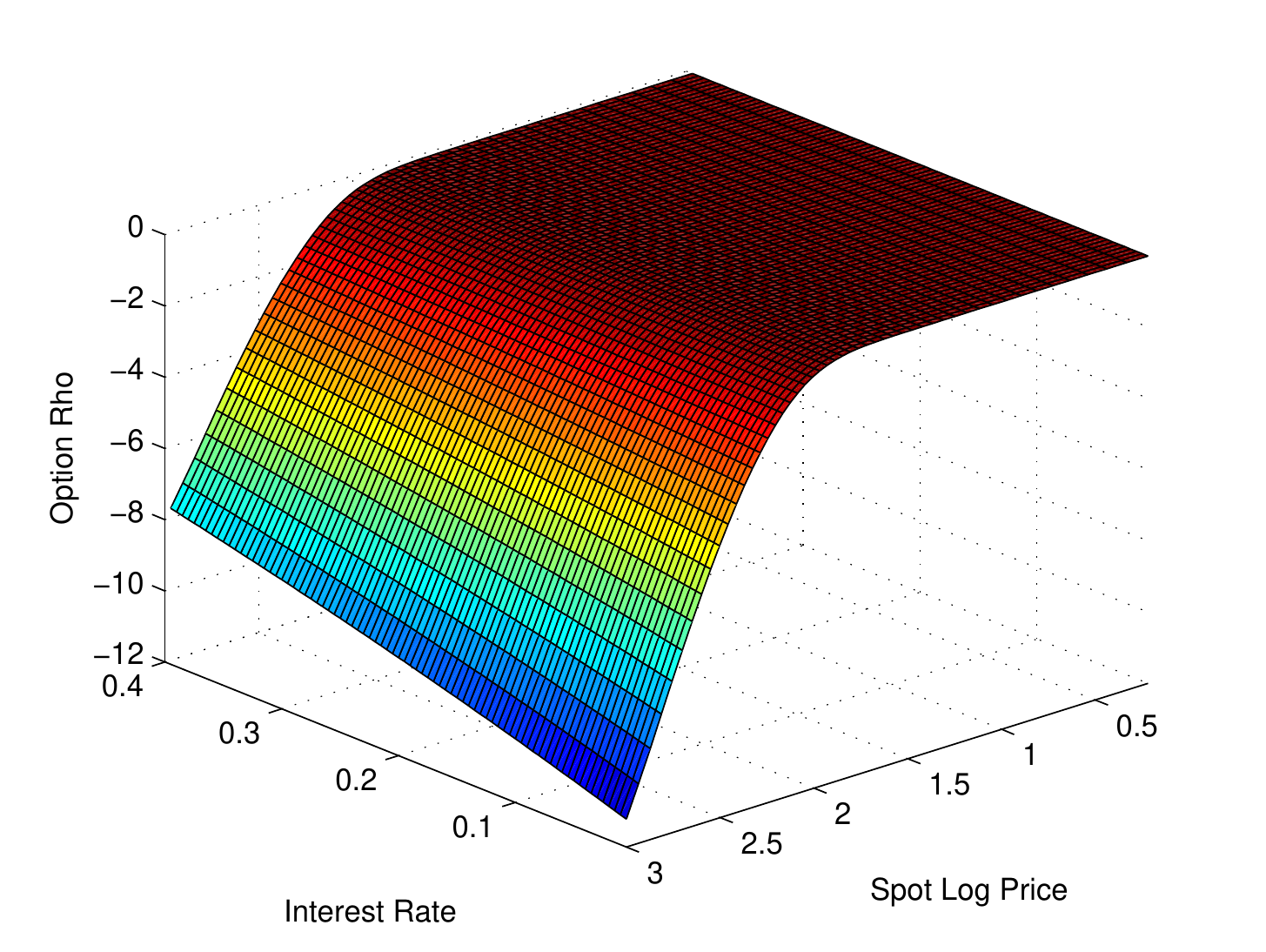}
  \caption{Option Rho}
  \label{fig:sub13}
\end{figure}

\noindent Figure \eqref{fig:sub11} shows that the option price is a decreasing monotonic function with respect to the interest rate. This result is confirmed by analyzing the first derivative of the option price with respect to the interest rate presented in figure \eqref{fig:sub13}, also known as Rho. When the option is out-of-the-money, Rho is approximately equal to zero. But as the option gets in-the-money, it is observed a negative slope. This result is counter-intuitive by considering that, under Black-Scholes model, the derivative is always positive. This discrepancy can be assessed by noting that, in equation \eqref{eq7}, the coefficient of $\frac{\partial \hat{V}}{\partial S} S$ is equal to $\, \left(q_S - \gamma_S \right)$ instead of $\left(r-\gamma_S\right)$. Given that under the $BK_{TC}$ model, $q_S$ is being modeled as a constant function, the positive sensitivity of the option to the interest rate is not observed. In order to match the expected behavior, an improvement of the modeling approach of the financing cost and its relationship with the interest rate has to be done.

\subsubsection{Transaction Costs}

\begin{figure}[h]
\centering
\begin{subfigure}{.5\textwidth}
  \centering
  \includegraphics[scale=0.5]{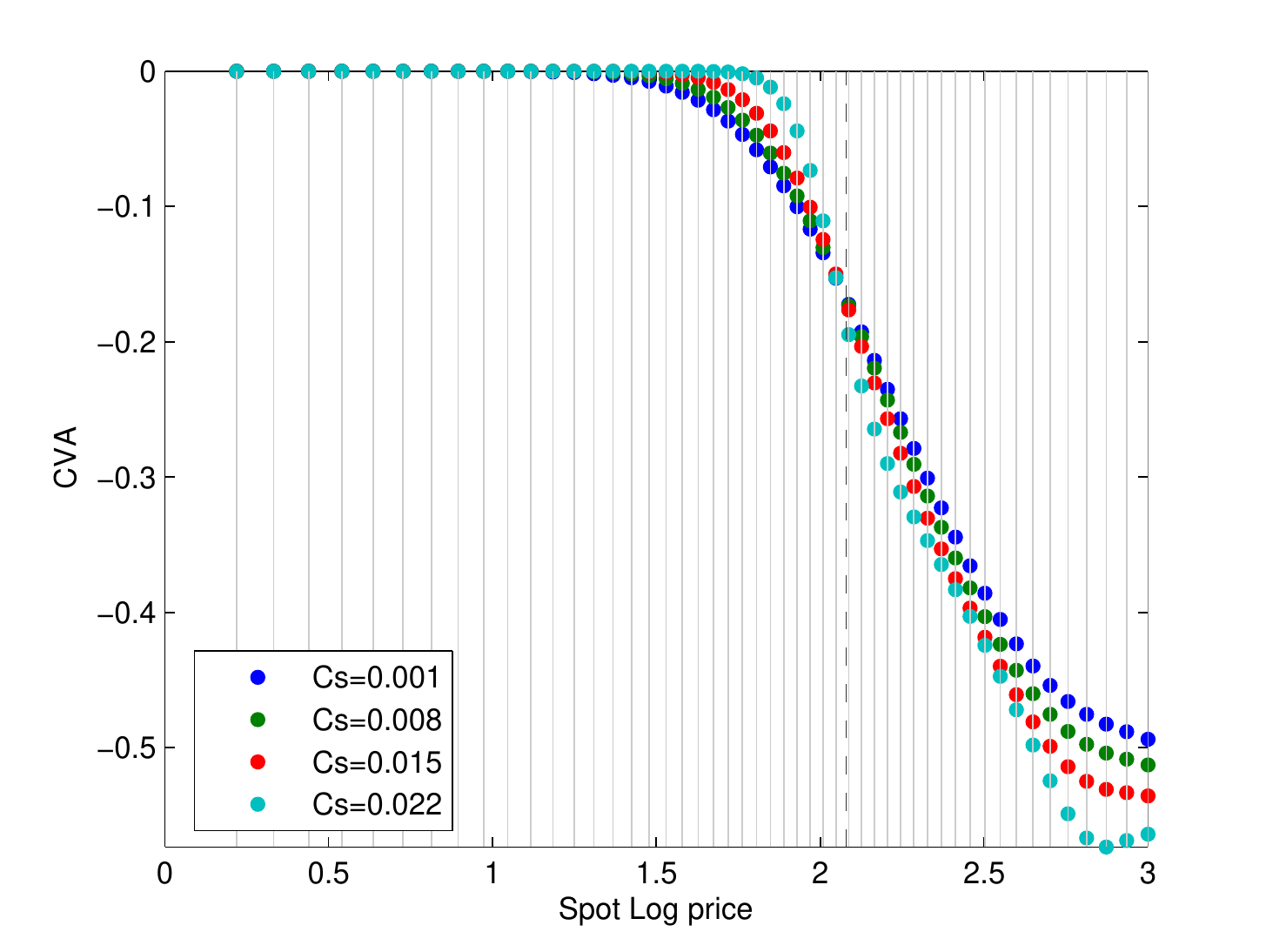}
  \caption{$BK_{TC}$}
  \label{fig:sub14}
\end{subfigure}%
\begin{subfigure}{.5\textwidth}
  \centering
  \includegraphics[scale=0.5]{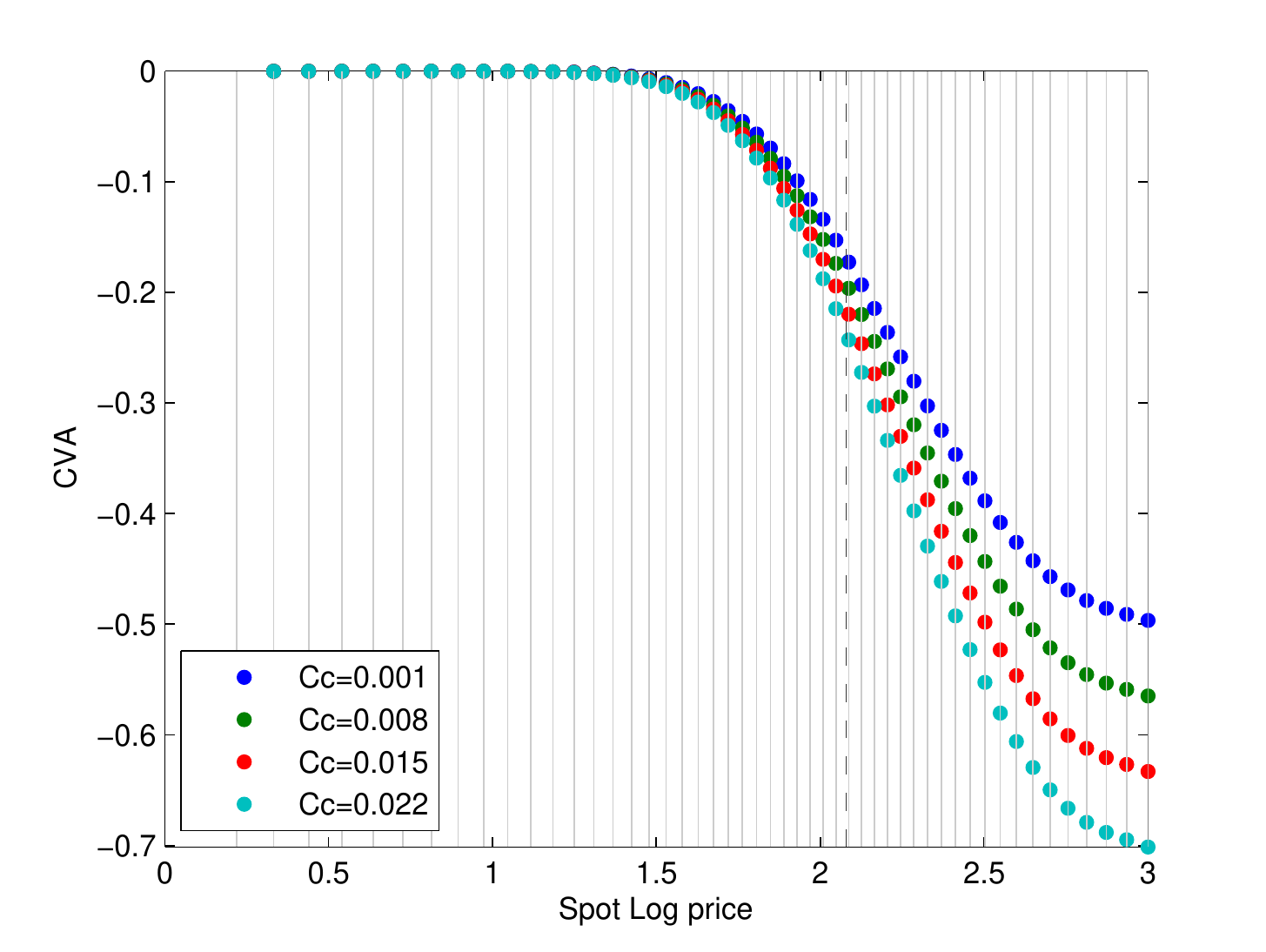}
  \caption{$BK_{TC}$}
  \label{fig:sub15}
\end{subfigure}
\caption{$r=0.05$, $q_S=0.05$, $\gamma_S=0.03$, $\sigma=0.1$, $S_f=0$, $\lambda_C = 0.01$, $R_C = 0.4$, $\lambda_B=0.05$, $R_B=0.4$, $C_B=0.001$, $dt=1/261$, $\Delta \tau=1/261$, $K=8$. $C_C=0.001$ for \eqref{fig:sub14} and $C_S =  0.002$ for \eqref{fig:sub15}.}
\label{fig:cost}
\end{figure}

\noindent Figure \eqref{fig:cost} presents the variation on the CVA due to changes in the transaction costs that arise of trading $\delta$ amount of shares $S$ and $\alpha_C$ amounts of bond $P_C$. By recalling equation \eqref{eq7} it can be noted that an increase in $C_S$ leads to a decrease in the modified volatility. We actually can assume that the modified volatility behaves similarly to the actual volatility, so that the analysis done in Section \ref{subsub:vol} can be applied. By considering the pattern showed in Figure \eqref{fig:cost} it can be seen that it is in line with the behavior of the CVA when varying the volatility in Figure \eqref{fig:sub14}. In both cases, the convexity changes near the strike value due to the same issues presented in the aforementioned section.\\

\noindent On the other side, the presence of $C_C$ in Equation \eqref{eq7} actually shows that larger costs generate a lower option price. Also, as transaction costs are multiplied by Delta, the gap widens as the option gets deeper in-the-money. This is the pattern that is observed in Figure \eqref{fig:sub15}.\\

\subsubsection{Recovery Rate and Hazard Rate}

\begin{figure}[hbtp]
\centering
\begin{subfigure}{.5\textwidth}
  \centering
  \includegraphics[scale=0.5]{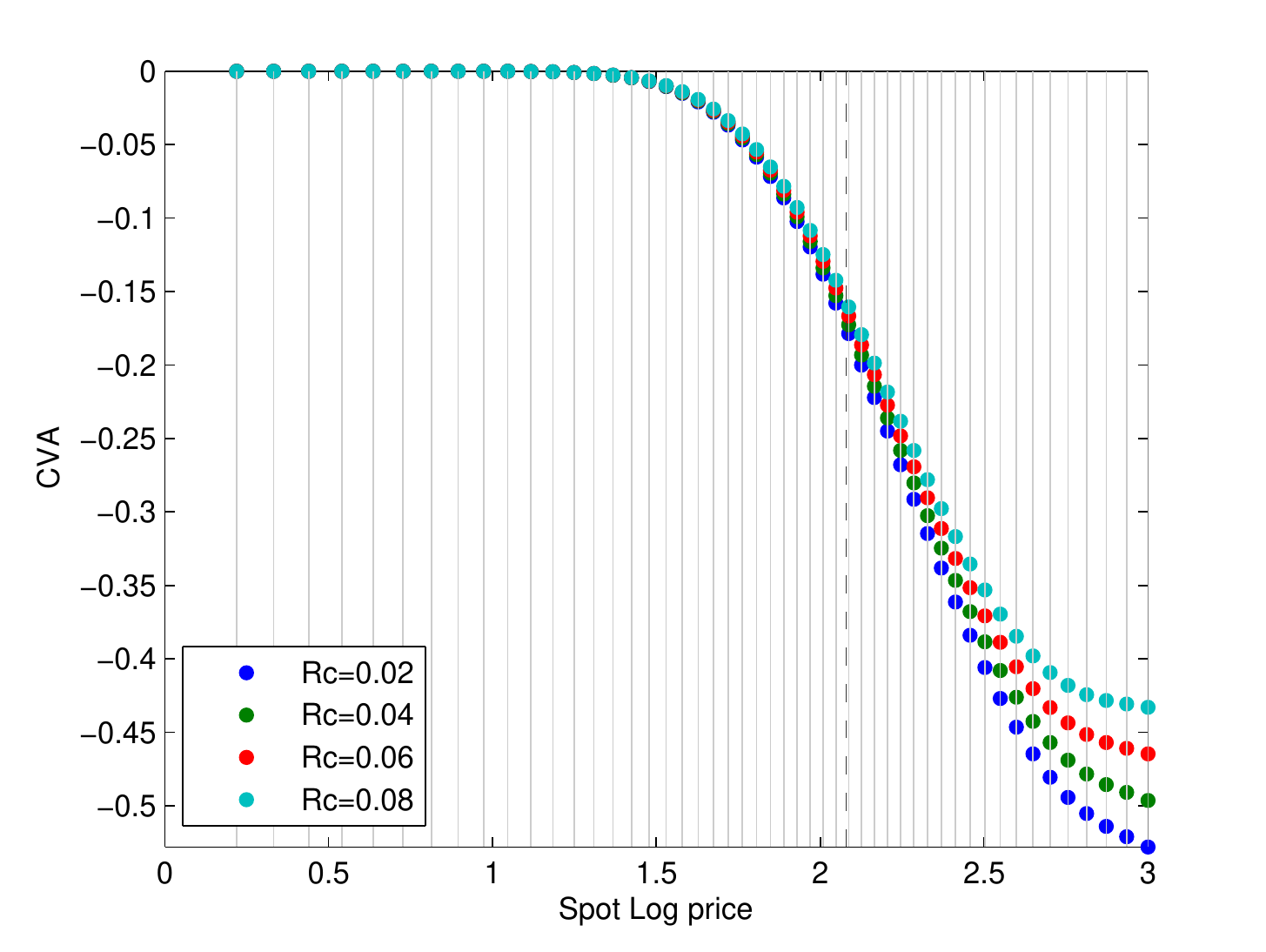}
  \caption{$BK_{TC}$}
  \label{fig:sub16}
\end{subfigure}%
\begin{subfigure}{.5\textwidth}
  \centering
  \includegraphics[scale=0.5]{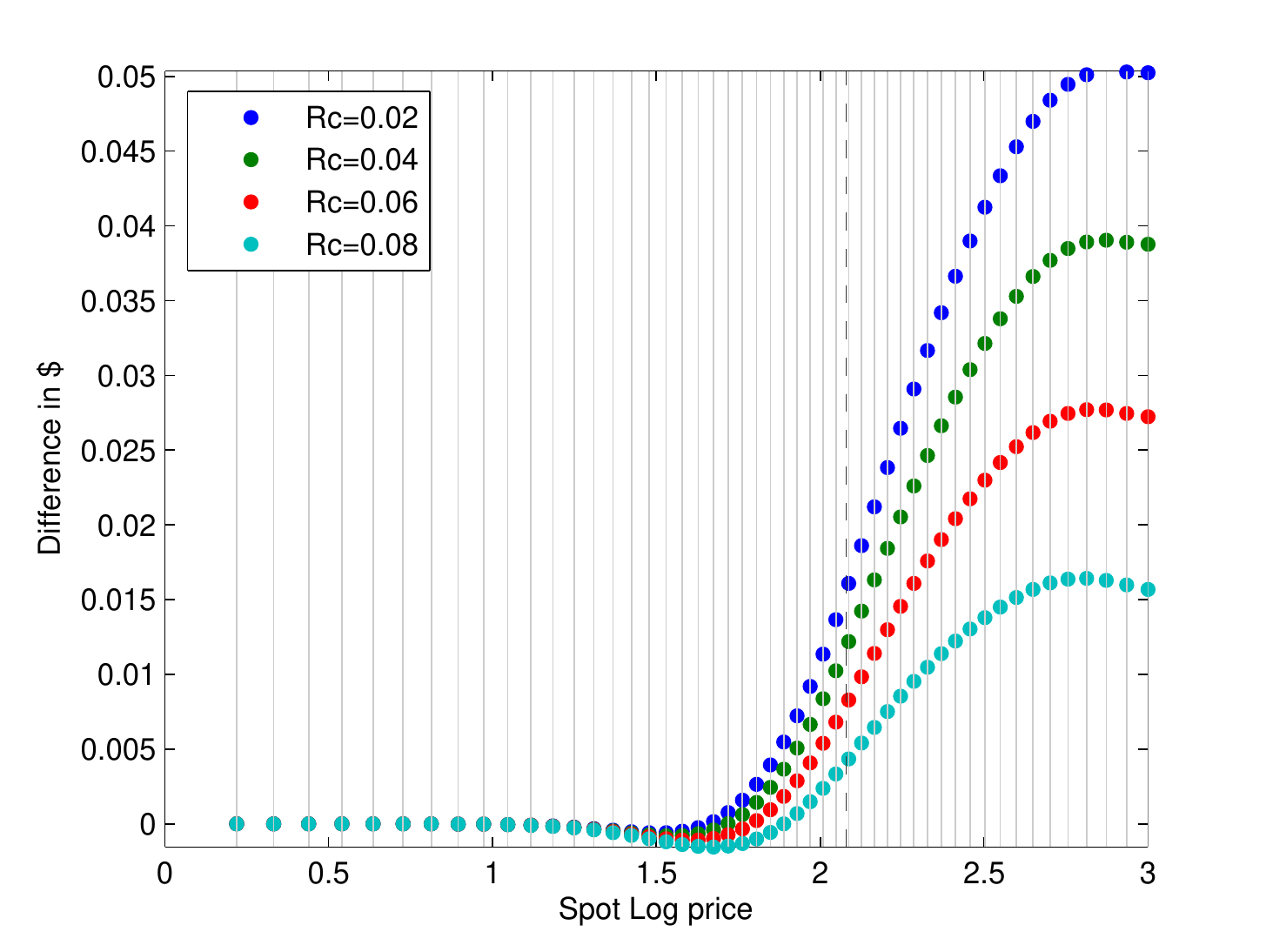}
  \caption{Difference between BK and $BK_{TC}$}
  \label{fig:sub17}
\end{subfigure}
\caption{$C_S =  0.002$, $r=0.05$, $q_S=0.05$, $\gamma_S=0.03$, $\sigma=0.1$, $S_f=0$, $\lambda_C = 0.01$, $\lambda_B=0.05$, $R_B=0.4$, $C_B=0.001$, $C_C=0.001$, $dt=1/261$, $\Delta \tau=1/261$, $K=8$}
\label{fig:Rc}
\end{figure}

\noindent In Figure \eqref{fig:Rc} the sensitivity of the CVA due to changes in the recovery rate of the counterparty bond is presented. Given that the recovery rate determines the amount of instrument that can be recovered in case of a default, the term $\left(1 - R_C \right)$ estimates the loss that would arise in case of default. It is expected that higher recovery rates imply lesser losses and then lesser CVA. Figure \eqref{fig:sub16} shows this monotonic relationship which is also in line with the behavior seen in Equation \eqref{eq7}.\\

\begin{figure}[hbtp]
\centering
\begin{subfigure}{.5\textwidth}
  \centering
  \includegraphics[scale=0.5]{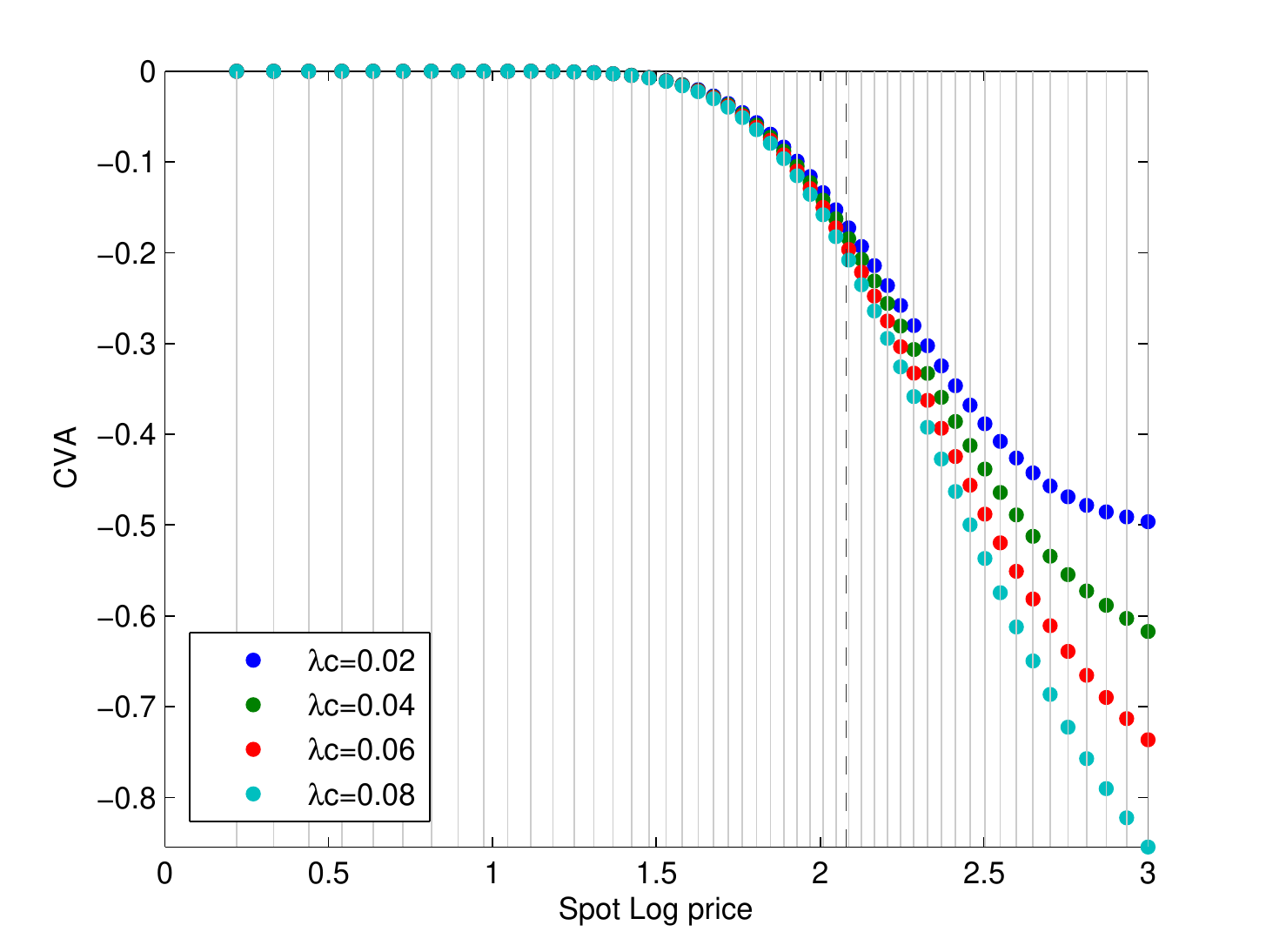}
  \caption{$BK_{TC}$}
  \label{fig:sub18}
\end{subfigure}%
\begin{subfigure}{.5\textwidth}
  \centering
  \includegraphics[scale=0.5]{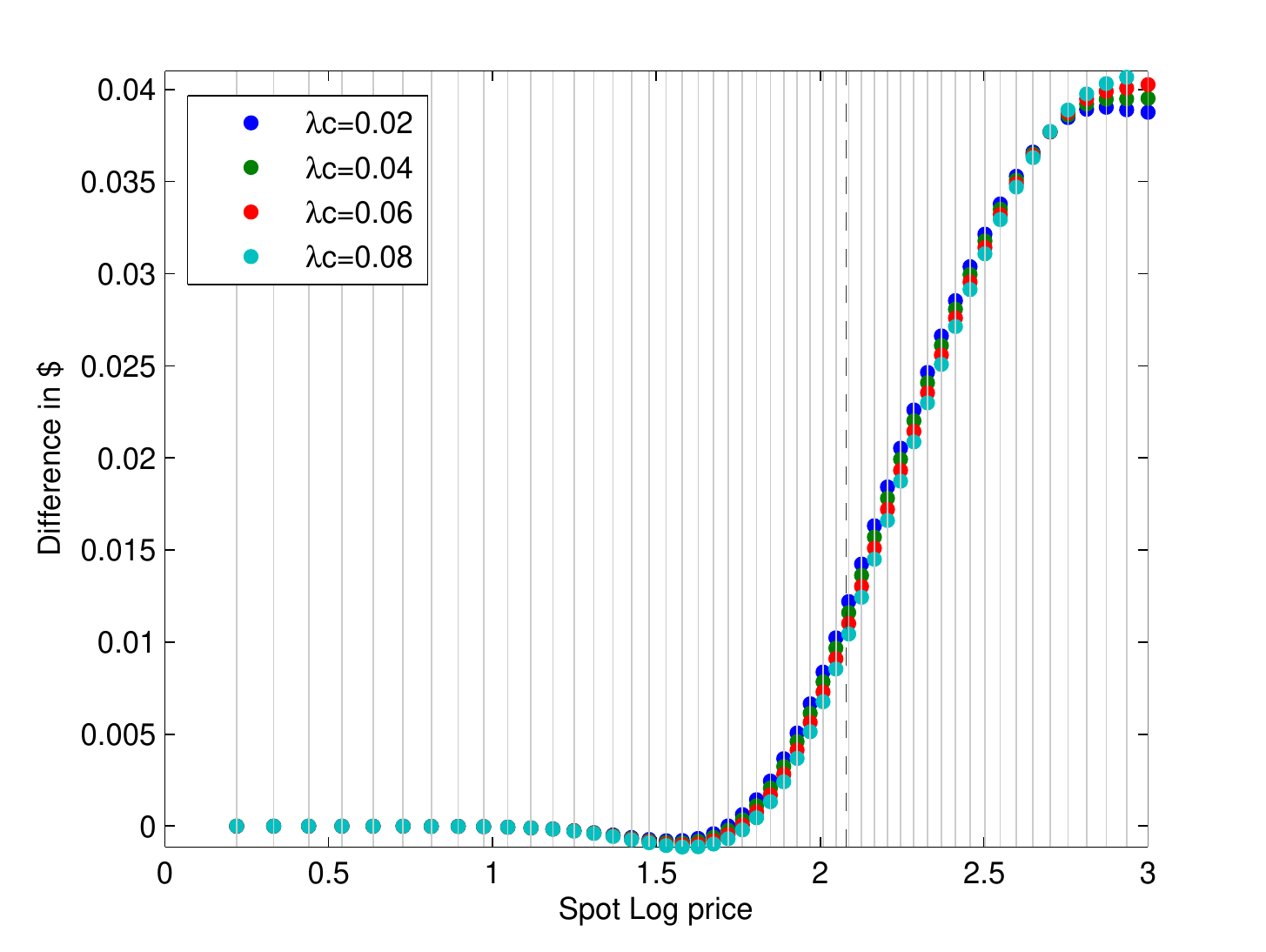}
  \caption{Difference between BK and $BK_{TC}$}
  \label{fig:sub19}
\end{subfigure}
\caption{$C_S =  0.002$, $r=0.05$, $q_S=0.05$, $\gamma_S=0.03$, $\sigma=0.1$, $S_f=0$, $\lambda_B=0.05$, $R_B=0.4$, $R_C = 0.4$, $C_B=0.001$, $C_C=0.001$, $dt=1/261$, $\Delta \tau=1/261$, $K=8$.}
\label{fig:lambdac}
\end{figure}

\noindent The hazard rate of the counterparty bond measures the likelihood that the bond will default at a certain point of time. Hence, if the hazard rate increases, the probability of default of the bond also increases. Then, it is expected to see a higher CVA value when deriving the option price. Figure \eqref{fig:sub18} presents the CVA value for different hazard rates where the expected behavior is noticed.\\

\section{Conclusion}

\noindent In this work we adapted the PDE formulation of CVA modeling originally proposed in \cite{burgard2011partial} to consider the existence of transaction costs when developing the hedging portfolio. We used the Leland approach to include constant transaction costs applied to trading the spot asset and both counterparty and seller bonds. We constructed a nonlinear PDE model which explains the behavior of an option price considering both counterparty risk and transaction costs. Further, we proved the existence of a solution by applying a fixed-point approach under a constraint in the volatility parameter. Numerical results showed that Delta and Gamma behave similarly as in a plain vanilla option but Vega and Rho presented differences in terms of the usual behavior. 
It is observed that the presence of transaction costs impact on the way that volatility and risk-free interest rate affects the option price. We also realized that the spot financing cost $q_S$ has to be linked with the risk-free interest rate to assure consistent results.\\

\noindent Further work should include the extensions proposed in \cite{burgard2011partial} such as modeling stochastic interest rates, stochastic hazard rates, proposing default time dependency or working with derivatives with more general payments. Nonetheless, the model can be extended by deriving not only a more complex transaction costs function such as the ones used in \cite{vsevvcovivc2016analysis} but also on studying the dynamics of the option price over a basket of assets as in \cite{amster2017pricing}.\\

\section{Acknowledgments}

This work
was partially supported by projects CONICET PIP 11220130100006CO 
and UBACyT 20020160100002BA.
\clearpage
\addcontentsline{toc}{chapter}{Bibliography}

\bibliography{bib}
\bibliographystyle{ieeetr}

\end{document}